\documentclass[conference]{IEEEtran}
\IEEEoverridecommandlockouts
\makeatletter
\def\ps@IEEEtitlepagestyle{%
\def\@oddfoot{\mycopyrightnotice}%
\def\@evenfoot{}%
}
\def\mycopyrightnotice{%
{978-1-6654-8045-1/22/\$31.00~\copyright~2022 IEEE\hfill} % Revise this line accordingly!
\gdef\mycopyrightnotice{}
}
\usepackage{cite}
\usepackage{amsmath,amssymb,amsfonts}
\usepackage{algorithmic}
\usepackage{graphicx}
\usepackage{textcomp}
\usepackage{xcolor}
\usepackage{url}
\usepackage{fixmath}
\usepackage{amsmath}
\usepackage{mathrsfs}
\usepackage{algorithm2e}
\usepackage{hyperref}
\usepackage{graphicx}
\usepackage{subfigure}
\usepackage{multirow}
\usepackage{float}
\usepackage{stfloats}
\usepackage{cancel}
\usepackage{booktabs}

\def\BibTeX{{\rm B\kern-.05em{\sc i\kern-.025em b}\kern-.08em
    T\kern-.1667em\lower.7ex\hbox{E}\kern-.125emX}}
\begin{document}

\title{Transition-Aware Multi-Activity Knowledge Tracing\\
%\thanks{This paper is based upon work supported by the National Science Foundation under Grant No. 2047500.}
}

\author{\IEEEauthorblockN{1\textsuperscript{st} Siqian Zhao}
\IEEEauthorblockA{\textit{Computer Science Department} \\
\textit{University at Albany, SUNY}\\
Albany, NY 12222,  USA \\
szhao2@albany.edu}
\and
\IEEEauthorblockN{2\textsuperscript{nd} Chunpai Wang}
\IEEEauthorblockA{\textit{Computer Science Department} \\
\textit{University at Albany, SUNY}\\
Albany, NY 12222, USA \\
cwang25@albany.edu}
\and
\IEEEauthorblockN{3\textsuperscript{rd} Shaghayegh Sahebi}
\IEEEauthorblockA{\textit{Computer Science Department} \\
\textit{University at Albany, SUNY}\\
Albany, NY 12222, USA \\
ssahebi@albany.edu}
}
% \and
% \and
% \and

%% Paper ID: BigD737   Paper Password: p7377891

\maketitle
% \IEEEpubidadjcol

\begin{abstract}
Accurate modeling of student knowledge is essential for large-scale online learning systems that are increasingly used for student training.
Knowledge tracing aims to model student knowledge state given the student's sequence of learning activities.
%As students interact with and learn from multiple types of learning materials, their knowledge is transferred between different material types.
Modern Knowledge tracing (KT) is usually formulated as a supervised sequence learning problem to predict students' future practice performance according to their past observed practice scores by summarizing student knowledge state as a set of evolving hidden variables.  % in a space that can be interpreted as learning material latent ``features'' or ``concepts''.
Because of this formulation, many current KT solutions are not fit for modeling student learning from non-assessed learning activities with no explicit feedback or score observation (e.g., watching video lectures that are not graded).
Additionally, these models cannot explicitly represent the dynamics of knowledge transfer among different learning activities,
%As students take on different learning activities, their gained knowledge transfers from one type of activity (e.g., watching a video lecture) to another (e.g., solving a problem). 
particularly between the assessed (e.g., quizzes) and non-assessed (e.g., video lectures) learning activities.
%However, current multi-type knowledge tracing methods neglect to investigate the dynamics of knowledge transfer between different material types.
%Transfer learning is a machine learning technique that has been successfully applied in other studies, such as recommender systems, to improve recommendations by transferring information between multiple domains.
In this paper, we propose Transition-Aware Multi-activity Knowledge Tracing (TAMKOT), which models knowledge transfer between learning materials, in addition to student knowledge, when students transition between and within assessed and non-assessed learning materials.
% TAMKOT is formulated as a deep recurrent semi-supervised multi-view learning model that explicitly learns knowledge transfer by activating and learning a set of knowledge transfer matrices, one for each transition type between student activities.
TAMKOT is formulated as a deep recurrent multi-activity learning model that explicitly learns knowledge transfer by activating and learning a set of knowledge transfer matrices, one for each transition type between student activities.
%To this end, we propose a  that allows students to switch freely between learning material types at any time step and can explicitly represent their knowledge transfer.
Accordingly, our model allows for representing each material type in a different yet transferrable latent space while maintaining student knowledge in a shared space.
We evaluate our model on three real-world publicly available datasets and demonstrate TAMKOT's capability in predicting student performance and modeling knowledge transfer. 
\end{abstract}


\section{Introduction}
\label{sec:intro}
The popularity and necessity of online education have increased the use of online education systems in recent years.
Given the amount of data produced as a result of student interactions, automatically understanding individual students' knowledge and learning process is essential for the success of such online education systems. 
Knowledge Tracing (KT) models aim to quantify students' state of knowledge at each point of the learning period.
Many modern KT models are formulated as a supervised sequence learning problem to predict students' future practice performance according to their past performances in learning activities~\cite{piech2015deep,drasgow1990item,yudelson2013individualized,lan2014sparse,piech2015deep,Khajah2014,Vie2019,zhang2017dynamic, sahebi2016tensor}.
Students learn from activities such as solving problems, taking tests, reviewing worked examples, and watching video lectures~\cite{nguyen2015effectiveness,romero2010educational} (multi-activity).
Among these activities, the ones that can be used to assess student performance, like problem-solving and test-taking, serve as relatively reliable measures of student knowledge.
Unlike these assessed activities, the non-assessed activities, such as reading worked examples, cannot provide explicit indicators of knowledge in students.
Until recently, modeling the assessed activity types has been the main focus of these supervised sequence learning KT methods, and 
these models overlook essential aspects of the student learning process by ignoring the non-assessed learning activities.

Indeed, research shows that non-assessed learning activities can help students to learn better~\cite{najar2014adaptive,agrawal2014mining}. 
But, the realization and attainment of the gained knowledge from the assessed and non-assessed learning materials can be different. 
%Particularly, one learning material type can be more or less helpful for student performance in different assessed activities.  
For example, Hou et al. conclude that practice activities are useful for student success in projects, but they do not help as much in exam preparation~\cite{hou2021drinking}. 
Instead, they show that reviewing practice quizzes could help for exam preparation.
In other words, the knowledge that is gained from one learning material type (e.g., video lectures) can be transferred to another (e.g., solving problems).
However, the dynamics and realization of this transfer depend on the transition order of learning materials. % the knowledge is being transferred from and to.
For example, consider a student who is learning about ``summation'' and ``multiplication'' concepts by watching video lectures and practicing problems.
Since the multiplication concept can be explained as an extension of summation, the student can learn multiplication easier from a video lecture after practicing summation problems.
Meaning that the summation knowledge that is gained by solving problems can be transferred to help achieve better multiplication knowledge using video lectures.
However, the reverse sequence may not be as helpful.
A student may not be able to solve summation problems just by watching multiplication videos if they do not have background knowledge in summation. 
\looseness=-1

As a result, explicitly modeling knowledge transfer between different learning material types, particularly both assessed and non-assessed ones, is essential to accurately understand student learning processes.
%Despite that student acquiring knowledge from multiple types of learning materials, modeling student knowledge from all the types of learning materials is still a problem under investigation.
%Some types of learning materials are lack of providing student learning feedback, such as watching video lectures, and participating in discussions. 
%The feedback is efficient since it can explicitly reflect the student's knowledge after interacting with learning materials. For example, if the student receives a high score after solving a question, it shows they most likely master the knowledge concept involved in this question. 
%Therefore, in the literature, previous knowledge tracing approaches mostly rely on assessed (gradable) student learning activities, 
%many existing state-of-art approaches had been shown to be successfully model the student knowledge. 
%In fact, previous research has shown that non-assessed learning materials can help students to learn better~\cite{najar2014adaptive, agrawal2011enriching, agrawal2014mining, hou2021drinking}. 
%For example, reading worked examples is considered as an effective learning strategy for students in solving problems~\cite{najar2014adaptive}. 
%Methods ignore the non-assessed learning additives lost the extensive knowledge information and are lack of modeling student knowledge transfer in the entire learning process.  
Recently, a handful of works have sought to model both assessed and non-assessed learning activities~\cite{zhao2020modeling,abdi2021open,wang2021learning}. 
% Very recently, a few researchers started to note the non-assessed learning activities in knowledge tracing. 
% In some approaches, the non-assessed learning activities are used as extra features when modeling the assessed learning activities~\cite{zhang2017incorporating, choi2020towards}.
% While these models successfully improve student performance prediction, they do not explicitly represent student knowledge gain and knowledge state when interacting with non-assessed learning materials.
%Other approaches explicitly model student knowledge state when working with both assessed and non-assessed learning material types~\cite{zhao2020modeling, wang2021learning, abdi2021open}.
% Zhao et al. proposed a multi-activity tensor factorization model that represents student knowledge and all learning material types in a shared latent concept space~\cite{zhao2020modeling};
% Abdi et al. presented a multi-activity factorization machine that relies on student activity count with each material to predict student outcome~\cite{abdi2021open}; 
% Wang et al. proposed a deep multi-activity memory-based model that can represent a fixed number of non-assessed activities between every two assessed ones~\cite{wang2021learning}.
However, none of these approaches model how student knowledge transfers from one learning activity type to another.
%these models ignore the sequential nature of student activities and rely on counting the number of precious success or failures~\cite{abdi2021open} or limit the non-assessed activity sequence length back on count  
%However, current multi-type knowledge tracing approaches face some limitations. 
%Both~\cite{abdi2021open} and~\cite{zhao2020modeling} do not explicitly model the knowledge transition between learning activities.
%~\cite{wang2021learning} can not model the knowledge transition from all non-assessed learning activities, and model knowledge transfer between different permutation of learning material types to be the same.
%We argue that the student transfer different knowledge between different permutation of learning material types. 
%The knowledge transfer from questions to lectures is different from lectures to questions. 
%Not all the knowledge acquired from lectures is useful for the student to solve the questions.
%Also, solving the questions may not help understand the lectures, especially when the lecture is for totally new concepts.
%The questions and lecture latent concept spaces should be different with some of the concepts being shared.
In this paper, we propose Transition-Aware Multi-activity Knowledge Tracing (TAMKOT) to explicitly model knowledge transfer every time a student transitions between different learning activity types. 
%To this end, 
TAMKOT models student knowledge states in a set of latent variables at every step in the student learning sequence. 
Every time a student transitions from one learning material to another, TAMKOT uses a transition-specific matrix to transfer the student's knowledge according to the type of involved learning activities.
Unlike previous KT models, our formulation allows for unlimited transitions between different learning activity types and does not limit sequence lengths for any of the material types.
This is realized via the simple, yet efficient, formulation of transition identifiers in TAMKOT that activate one transition-specific matrix at a time.
%Furthermore, TAMKOT learns a distinct latent representation for each learning material that can be used to find conceptual similarities in them.
Our model provides the flexibility for different material types to have different latent representation spaces that are mapped to a shared student knowledge space.

%Knowledge transfer between learning material concepts is modeled via transition-specific transfer weight parameters that are activated according to transition identifiers.
%The transition identifiers specify what transition has happened between which learning material types in the student sequence.

%The simple, yet effective, formulation of TAMKOT doesn't need alignment of task sequences and allows for any number of transitions at any point during the sequence.

We evaluate TAMKOT on three real-world datasets. 
The experiments show that TAMKOT performs significantly better than state-of-the-art supervised knowledge tracing models in predicting student performance.
Furthermore, despite its simplicity, TAMKOT performs better than the existing multi-activity knowledge tracing models in datasets with granular learning materials. %in terms of accuracy and computation time, 
More importantly, our analysis demonstrates that knowledge transfer can be different depending on the transition order between learning material types, especially in complex learning materials.
Finally, we showcase the interpretability of the learned student knowledge states. %, and the meaningful learning material similarities discovered based on their recovered latent concepts. 
%The knowledge transferred by switching from watching video lectures to solving problems is not necessarily the same as the knowledge transfer from problems to video lectures.
%This transfer varies based on the concept that is being transferred.
%The significance of this difference varies in datasets with different learning material structures and complexity.
%Particularly, this difference is more significant in complex learning materials 

% In summary, our contributions to this paper are:
% \begin{itemize}
%     \item proposing a multi-type deep learning-based knowledge tracing model (TAMKOT) that can trace knowledge state from both assessed and non-assessed learning activities, can model different latent concept spaces for different learning material types.
%     \item conducting the experiments on three real-world datasets that indicate TAMKOT is similar to, or better than baseline methods for student performance prediction.
%     \item  designing the analyses that demonstrate the knowledge transfer difference between various permutations of learning material types, show TAMKOT learns meaningful knowledge states, and show TAMKOT captures the interpretable latent learning material concepts.
% \end{itemize}

\section{Related Work}
\label{sec:related}
In general, our work relates to the problem of KT, which quantifies a student's knowledge state at each learning time step based on the student's sequence of previous learning activities.
% First, pioneers started to address the problem of KT based on regression and traditional probabilistic models~\cite{doan2019rank}.
% First, regression-based KT models~\cite{drasgow1990item, cen2006learning, cen2008comparing, pavlik2009performance, lan2014sparse} utilized logistic regression to capture the relationship between student performance and several factors of student and question.
This quantification of student knowledge is usually defined over a set of knowledge components or concepts of the learning materials.
For example, an Algebra problem can include knowledge components related to the Summation and Multiplication concepts and the students learn those concepts by attempting that problem.
Traditionally, many KT models, such as Bayesian knowledge tracing (BKT)~\cite{corbett1994knowledge}, and regression-based KT models~\cite{drasgow1990item, cen2006learning, cen2008comparing, pavlik2009performance}, rely on a \textit{predefined mapping} between the learning materials and knowledge components as their input.
%These approaches use the predefined mapping to predict the probability that a student will correctly answer a question with the labeled KCs.
This mapping could be achieved by human experts labeling the learning materials.  
%Accurate labeling of knowledge components needs human experts to assign concepts/topics/skills to each learning material.
% \textcolor{blue}{In this pioneering work, the knowledge components of each question are predefined and provided.
% These approaches use the predefined knowledge components to predict the probability that the student will correctly answer the question with labeled knowledge components, such as bayesian knowledge tracing (BKT)~\cite{corbett1994knowledge}, and regression-based KT models~\cite{drasgow1990item, cen2006learning, cen2008comparing, pavlik2009performance, lan2014sparse}.
% }
However, while labeling the concepts for each learning material can be done in smaller educational systems, it is not feasible for large-scale systems that present thousands of learning materials to students.

Later, some KT models were designed to avoid requiring such a predefined mapping.
These KT models learn the underlying \textit{latent concepts} presented in the learning materials for this purpose~\cite{thai2010factorization,sahebi2016tensor,zhang2017dynamic}.
This includes state-of-the-art deep learning-based KT methods.
For example, Zhang et al. proposed dynamic key-value memory network (DKVMN) that uses a variant of memory augmented neural networks for modeling the underlying learning material concepts and student knowledge state~\cite{zhang2017dynamic}.
% Dynamic key-value memory networks (DKVMN) proposed in~\cite{zhang2017dynamic} uses a variant of MANN with the attention mechanism for better modeling underlying learning material concepts and student knowledge state through the external memory.
% Pandey et al. introduces SAKT, an attention-based method that leverages self-attention mechanism to model interdependencies among interactions at each time step~\cite{pandey2019self}.
DKT, the first deep learning-based KT method, %learns latent concept representations of learning materials and 
utilizes the Recurrent Neural Networks (RNN) to model student knowledge state~\cite{piech2015deep}.
Later, attention mechanisms have been used to improve KT models.
For instance, Pandey et al. introduced SAKT, an attention-based method that models the interdependencies among student interactions at each attempt~\cite{pandey2019self}.
% In RKT~\cite{pandey2020rkt} and MF-DAKT~\cite{zhang2021multi}, Pandey et al. and Zhang et al. proposed to leverage contextual information with attention mechanism for improving KT.
As another example, AKT is an attentive method that learns context-aware question and student performance representations and utilizes the attention mechanism to model student knowledge state~\cite{ghosh2020context}.

However, all the approaches mentioned above %are supervised KT models that 
ignore students' non-assessed learning activities and only consider the assessed learning materials.
In a few recent approaches, non-assessed learning activities are used as extra features when modeling assessed learning activities~\cite{zhang2017incorporating, choi2020towards}.
While these models successfully improve student performance prediction, they do not explicitly represent student knowledge gain and knowledge state when interacting with non-assessed learning materials.
To the best of our knowledge, MA-Elo~\cite{abdi2021open}, MA-FM~\cite{abdi2022learner}, MVKM~\cite{zhao2020modeling}, and DMKT~\cite{wang2021learning} are the only \textit{multi-activity KT models} that explicitly model student knowledge state from multiple types of learning activities. %, out of which only MVKM and DMKT can be compared to our work.
% Abdi et al. proposed MA-Elo in~\cite{abdi2021open}, it is a method that captures the representation of a student's knowledge state based on their higher-order learning activities such as creating materials and rating the quality of materials.
MA-Elo is a multivariate Elo-based~\cite{wauters2010monitoring} learner model that adjusts student knowledge state in a predefined set of knowledge concepts according to the difference between the student's predicted and observed next activity~\cite{abdi2021open}.
%Abdi et al. also proposed another multi-activity method, 
MA-FM is based on factorization machines and models student knowledge state in predefined knowledge concepts by a weighted count of the student's previous successes, failures, and activities with different learning material types~\cite{abdi2022learner}.
Both MA-FM and MA-Elo require the predefined mapping between the learning materials and concepts, and cannot be used in systems that do not include this information.  %an expert-defined set of knowledge components or concepts for each learning material as an input. 
% While predefining the concepts of each learning material can be done in smaller educational systems, it is not feasible for large-scale systems that present thousands of learning materials to students. 
%In this paper, our goal is to model student knowledge without requiring such mapping. % learn learning material representations to discover latent knowledge concepts.
% Zhao et al. presents MVKM~\cite{zhao2020modeling} that models student learning activities of different learning material types as separate tensors and assumes that their latent concepts are shared.
% It uses tensor factorization to capture the latent students’ features, student knowledge at each time step and learning material latent concepts.
MVKM and DMKT are the only multi-activity models that do not require this information.
MVKM models student learning activities of different material types as separate tensors and uses tensor factorization to capture latent students’ features, student knowledge, and learning material latent concepts~\cite{zhao2020modeling}.
DMKT extends a KT model for assessed activities, DKVMN~\cite{zhang2017dynamic}, and uses key-value memory networks to model student knowledge gain over both assessed and non-assessed learning activities~\cite{wang2021learning}.
Since DMKT has a fixed architecture that does not adjust to the learning activity types, it can only model a fixed number of non-assessed activities between every two assessed ones.
None of the above methods explicitly represent the dynamics of knowledge transfer among different learning activities. 
Moreover, MVKM and DMKT assume the exact same latent concept spaces for different types of materials, while different types of learning materials could have different knowledge concepts. %, some of them are shared.

\section{Problem Formulation}
\label{sec:problem_formu}
%As we mentioned that 
%In KT, % which only consider student interactions with the assessed learning material type, % usually questions.
%Accordingly, knowledge tracing is formulated 

KT is usually evaluated by the task of student performance prediction, where students' upcoming performances are predicted, given their past learning activities.
KT methods predominantly focus on assessed learning activities as students' past activities. % only student performance in the assessed learning materials, such as in the questions and the problems, are considered . 
Specifically, a student's interaction at each time step $t$ is denoted as $(q_t, r_t)$, where $q_t$ represents the assessed learning material (e.g., problem) that the student interacts with at the time step $t$, and $r_t$ denotes the student's performance (e.g, score, correctness, or grade) in $q_t$.
Given the previous performance records of a student as $\{(q_1, r_1), \dots, (q_{t}, r_{t})\}$, KT aims to predict the student's future performance $r_{t+1}$ in a problem $q_{t+1}$ at time step $t+1$.

Our goal is to trace students' knowledge at each time step $t$ as they learn from both assessed and non-assessed learning material types, explicitly model the knowledge transfer from each learning material type to another, and predict student performance on future assessed learning materials.
Additionally, we would like the model to represent unlimited student transitions between different activity types with no particular order.
Without loss of generality, assume an education system with one assessed learning material type (e.g., problems) and one non-assessed learning material type (e.g., video lectures).
%Without loss of generality, assume an education system has two types of learning materials, with $S$ students interacting with one set of $Q$ assessed learning materials (e.g., problems) and one set of $L$ non-assessed learning materials (e.g., video lectures).}
%\textcolor{red}{We denote a student's activities at each time step $t$ as a tuple $\left\langle(q_t, r_t), l_t, d_t)\right\rangle$, where $(q_t, r_t)$ represents student interacts with problem $q_t$ at time step $t$ and their performance is $r_t$ (e.g., score, correctness), $l_t$ represents the video lecture that the student watches at time step $t$, and $d_t$ is used to denotes the learning material type that the student interacts with at time step $t$.}
%\textcolor{red}{In the case of having two learning material types, $d_t \in \{0, 1\}$, where $0$ represents the assessed learning material type and $1$ represents the non-assessed type. 
%Since each student only interacts with one learning material at each time step $t$, we formulate the problem to only consider one learning material at a time, that is associated with the corresponding $d_t$.
%To this end, we represent a student's whole trajectory of activities with different types of learning materials as a sequence of tuples $\{\left\langle(q_1, r_1), l_1, d_1)\right\rangle, ..., \left\langle(q_{t}, r_{t}), l_{t}, d_{t})\right\rangle\}$, and use this sequence to predict student's future performance $r_{t+1}$ in a problem $q_{t+1}$ at time step $t+1$.}
Each student only interacts with one learning material, either an assessed or a non-assessed one, at each time step $t$.
We represent student activity type using an indicator $d_t \in \{0, 1\}$, where $0$ represents the assessed learning material type, and $1$ represents the non-assessed type. 
We also denote a student's activity at each time step $t$ as a tuple $\langle i_t, d_t\rangle$, where
$$i_t = \begin{cases}(q_t, r_t) & \text{if } d_t = 0 \\ l_t & \text{if } d_t = 1 \end{cases}$$  %Since each student only interacts with one learning material at each time step $t$, we formulate the problem to only consider one learning material at a time, that is associated with the corresponding $d_t$.
Here $(q_t, r_t)$ shows that the student interacts with the problem $q_t$ at time step $t$ with performance $r_t$, and $l_t$ represents the video lecture that the student watches at time step $t$.
This formulation allows us to represent student learning activities with both learning material types.
%For example, if $d_t = 0$, we obtain $\langle i_t, r_t\rangle = \langle (q_t, r_t), d_t\rangle$ that represents the student interacts with problem $q_t$ at time step $t$ and their performance is $r_t$ (e.g., score, correctness).
%Correspondingly, if $d_t = 1$, meaning $\langle i_t, r_t\rangle = \langle l_t, d_t\rangle$, it represents the student watches the video lecture $l_t$ at time step $t$.
Eventually, we represent a student's whole trajectory of activities with different learning materials types as a sequence of tuples $\{\left\langle i_1, d_1\right\rangle, ..., \left\langle i_t, d_{t}\right\rangle\}$.
%and use this sequence to predict the student's future performance $r_{t+1}$ in a problem $q_{t+1}$ at time step $t+1$.

To achieve our goal of predicting student performance, given their assessed and non-assessed learning activity history, we assume that
students gain knowledge in a set of latent concepts or topics that are presented in learning materials. 
However, the realization of student knowledge can vary in different material types.
We also assume that the knowledge gained using one learning material can be transferred to another learning material when students switch between them. 
\vspace{-5pt}

\section{TAMKOT Model}
\label{sec:mvrbtf}
\begin{figure*}[!ht]
    % \vspace{-5pt}
    \centering
    \subfigure{
    % \includegraphics[width=0.7\textwidth] {figures/model.pdf}}
        % \subfigure{
    \includegraphics[width=0.725\textwidth]{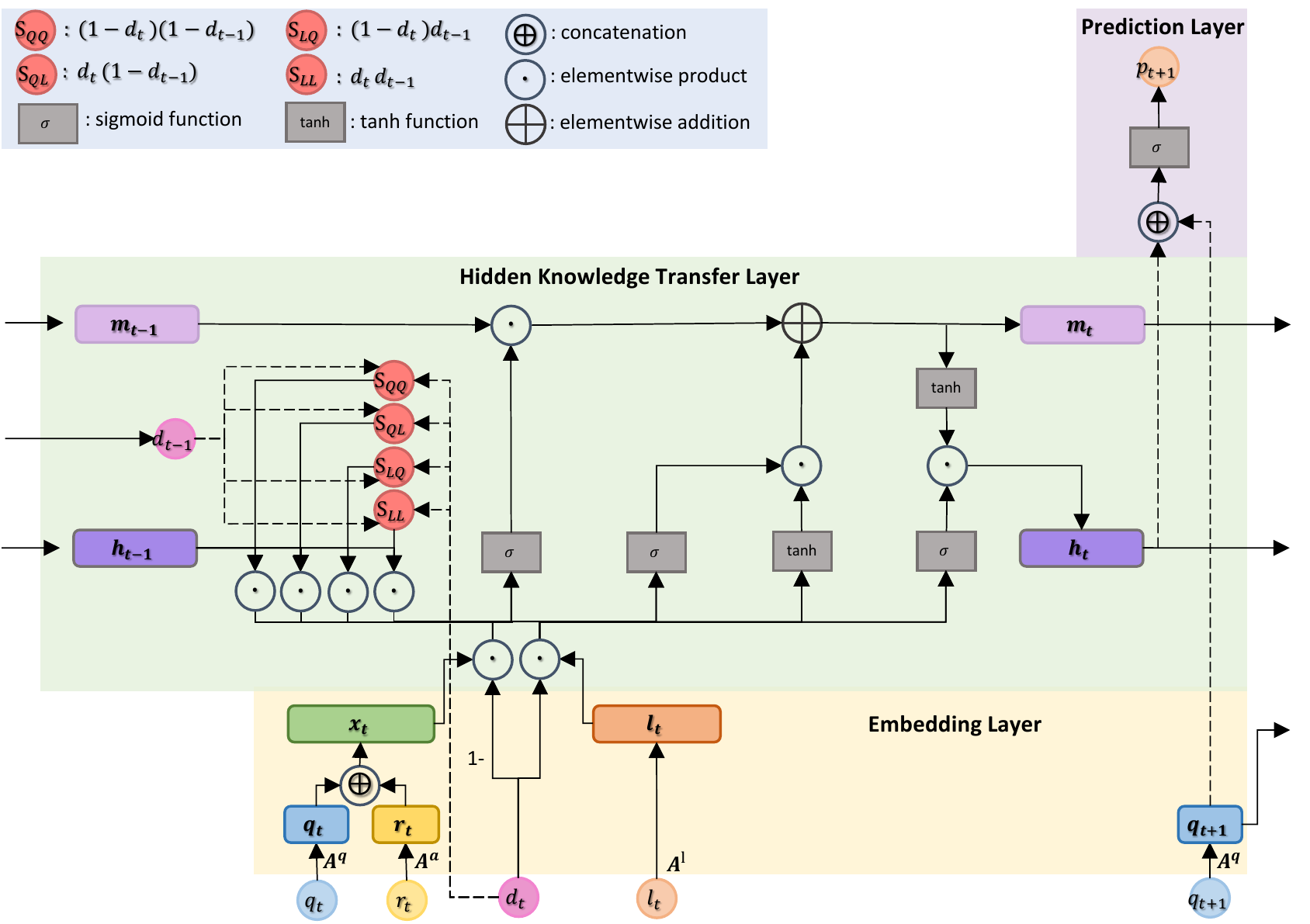}}
    
    % \vspace{-5pt}
    \caption{The TAMKOT model architecture. The solid lines and dashed lines are the same. Different line types are used to keep clarity between the lines that fall over/cross each other.}
    \label{fig:model}
    \vspace{-5pt}
\end{figure*}

In this section, we introduce our model Transition-Aware Multi-view Knowledge Tracing (TAMKOT). %, a deep recurrent multi-task learning model according to the aforementioned assumptions. 
% We build TAMKOT into three layers: (1) the embedding layer that maps each learning material and student's performance if available to the latent embedding space, (2) the hidden layer to model and transfer the knowledge between assessed and non-assessed interactions at each time step, and (3) the prediction layer to predict student's performance on an upcoming assessed learning material.
We build TAMKOT into three layers: (1) the embedding layer that maps each learning activity to the latent embedding space, (2) the hidden layer to model and transfer the knowledge between assessed and non-assessed interactions at each time step, and (3) the prediction layer to predict student's performance on an upcoming assessed learning material.
We formulate TAMKOT by building a transition-aware multi-activity component on top of LSTM~\cite{hochreiter1997long}.
%However, TAMKOT can be built on other deep recurrent models as well.
An overview of TAMKOT's architecture is presented in Figure \ref{fig:model}. In the following, we introduce the details of each layer. \looseness=-1

\textbf{Notations.} We use lowercase letters to denote scalars, e.g., $q_t$; boldface capital letters for matrices, e.g., $\mathbold{A}^q$; and boldface lowercase letters for vectors, e.g., $\mathbold{q}_t$.

\subsection{Embedding Layer}\label{sec:embedding_layer}
% The few previous multi-type knowledge tracing methods model both assessed and non-assessed learning materials into the same latent space or in the same embedding size~\cite{zhao2020modeling, wang2021learning}. 
% Our model TAMKOT takes each student interaction $\left\langle(q_t, r_t), l_t, d_t)\right\rangle$ in learning sequence as the input. 
% The goal of the embedding layer is to map each learning material ($q_t$ and $l_t$) and student response ($r_t$) into a 
% lower-dimensional 
% latent space. 
The goal of this layer is to learn the embedding vector of each learning activity $\left\langle i_t, d_{t}\right\rangle$ as the input to hidden knowledge transfer layer for estimating the student's knowledge hidden state $\mathbold{h}_t$, using the latent representation of its learning material ($q_t$ and $l_t$) and student response ($r_t$).
% lower-dimensional 
% latent space. .}
The few existing multi-activity KT methods model both assessed and non-assessed learning materials in the same latent space with the same dimensionality~\cite{zhao2020modeling, wang2021learning}. 
%This limitation would result in the same  assessed and non-assessed learning materials to have the same embedding size.
Unlike these works, we assume that the assessed and non-assessed learning materials can have different latent spaces. 
%In reality, different types of learning materials could have different latent spaces with some concepts shared.
This allows TAMKOT to be flexible in having more (or less) fine-grained representation for each learning material type. %, and in modeling non-shared latent concepts in different material types.
%Questions and lectures do not always share all the concepts. Some concepts in the lectures are not included in the questions. 
% To simplify the illustration, let us assume that the assessed and non-assessed types of learning materials are problems and video lectures, respectively.
%Let us assume the assessed learning material types is problem and non-assessed learning material type is video lectures.
Having problems as assessed and video lectures as non-assessed learning materials,
we first map all problems into the problem latent space
and video lectures into the video latent space and achieve their underlying latent concepts matrices $\mathbold{A}^q \in \mathbb{R}^{Q \times d_q}$ (for problems) and $\mathbold{A}^l \in \mathbb{R}^{L \times d_l}$ (for video lectures).
Here, $Q$ and $L$ are the number of questions and video lectures, and $d_q$ and $d_l$ are latent concept sizes for questions and video lectures respectively.
% \textcolor{blue}{we first map all problems and video lectures into two latent concept spaces using embedding matrices $\mathbold{A}^q \in \mathbb{R}^{Q \times d_q}$ and $\mathbold{A}^l \in \mathbb{R}^{L \times d_l}$, respectively, where $Q$ and $L$  are the number of questions and video lectures, $d_q$ is the problem latent concept size, and $d_l$ is the lecture latent concept size.}
% --For each assessed learning material, we use another embedding matrix $\mathbold{A}^r$ that maps student performance into the latent space. % with the embedding size $d_r$. % \in \mathbb{R}^{2 \times d_r}
For student performance $r_t$ in assessed learning materials, we use another embedding matrix $\mathbold{A}^r$ that maps student performance into the latent space.
When modeling binary student performance outcomes (e.g., success or failure in solving problems), $\mathbold{A}^r \in \mathbb{R}^{2 \times d_r}$, where $d_r$ is the performance embedding size. 
% For example, when $r_t = 0$, the embedding vector is the first row of $\mathbold{A}^r$.
%The dimension $2 \times d_r$ is because usually, a student's performance on a question is the correctness of the answer on the question, which has two outcomes success or failure. 
For modeling numerical performance outcomes (e.g., exam scores between $0$ and $1$), we use a linear mapping $f(r_t) = r_t \mathbold{A}^r$ that maps the numerical performance into higher dimension, and $\mathbold{A}^r \in \mathbb{R}^{d_r}$. 
% For example, when $r_t = 0.1$, the embedding vector of $r_t$ is equal to $0.1\mathbold{A}^r$.

At each time step $t$, TAMKOT looks up latent learning material and student performance representations for the learning activity $\left\langle i_t, d_{t}\right\rangle$ to create its embedding vector.
% Here $[\mathbold{q}_t \oplus \mathbold{r}_t]$ is the embedding vector for the assessed material activity, we look up question $q_t$'s embedding $\mathbold{q}_t \in \mathbb{R}^{d_q}$ and student performance outcome $r_t$'s embedding $\mathbold{r}_t \in \mathbb{R}^{d_r}$. And we concatenate the problem embedding vector $\mathbold{q}_t$ and student performance embedding vector $\mathbold{r}_t$ as $\mathbold{x}_t = [\mathbold{q}_t \oplus \mathbold{r}_t]$ as the activity's embedding.
For the problem activity $i_t = (q_t, r_t)$, it looks up latent representation $\mathbold{q}_t \in \mathbb{R}^{d_q}$ for the problem $q_t$, and $\mathbold{r}_t \in \mathbb{R}^{d_r}$ for the student performance outcome $r_t$. 
It then concatenates them  %$\mathbold{q}_t$ and $\mathbold{r}_t$ 
as $\mathbold{x}_t = [\mathbold{q}_t \oplus \mathbold{r}_t]$ to create the activity $i_t$'s embedding.
% And we concatenate the problem embedding vector $\mathbold{q}_t$ and student performance embedding vector $\mathbold{r}_t$ as $\mathbold{x}_t = [\mathbold{q}_t \oplus \mathbold{r}_t]$ as the activity's embedding.
For the video lecture activity $i_t = l_t$, it looks up lecture $l_t$'s latent representation $\mathbold{l}_t \in \mathbb{R}^{d_l}$ as activity $i_t$'s embedding.

% \textcolor{blue}{At each time step $t$, TAMKOT look up latent learning material and student performance representations to convert learning activity $\left\langle i_t, d_{t}\right\rangle$ to a embedding vector $\mathbold{i_t}$:
% $$\mathbold{i_t} = \begin{cases}[\mathbold{q}_t \oplus \mathbold{r}_t] & \text{if } d_t = 0 \\ \mathbold{l}_t & \text{if } d_t = 1 \end{cases}$$
% Here $[\mathbold{q}_t \oplus \mathbold{r}_t]$ is the embedding vector for the question activity, where $\mathbold{q}_t \in \mathbb{R}^{d_q}$ and $\mathbold{r}_t \in \mathbb{R}^{d_r}$ is latent representation of question $q_t$ and student performance outcome $r_t$, and $\oplus$ denotes the concatenation.
% % $\mathbold{l}_t$ represents the embedding vector of the video lecture activity that is the latent representation of video lecture $l_t$.}
% $\mathbold{l}_t$ is video lecture $l_t$'s latent representation and we use it as the embedding vector of the video lecture activity.}

% and use the concatenation as follow to be the input for the knowledge transfer unit: 
% For the assessed type, the embedding vectors $\mathbold{q}_t$ and $\mathbold{r}_t$ are concatenated as follows, to be used as input for the knowledge transfer unit:
% $$\mathbold{x}_t = [\mathbold{q}_t \oplus \mathbold{r}_t]$$
%where $\oplus$ denotes concatenation operation. % $\mathbold{x}_t$ become the new embedding vector for assignment and performance together.
%We use the concatenation $\mathbold{x}_t$ for assessed activities to be the input of the knowledge transfer unit.

\subsection{Hidden Knowledge Transfer Layer}
% \vspace{-5pt}
%Students cross-interact with assessed and non-assessed learning materials, their knowledge is transferred when they switch from one learning material type to another one. 
The hidden knowledge transfer layer is designed to represent the student' knowledge state $\mathbold{h}_t$ and learn knowledge transfer while they are freely interacting with and transitioning between assessed and non-assessed learning material types.
%In this hidden knowledge transfer layer, we model the knowledge transfer between student's assessed and non-assessed learning interactions. We base our model TAMKOT upon the Long Short-Term Memory (LSTM) technique.
%This layer is inspired by recurrent neural networks, particularly Long Short-Term Memory (LSTM)~\cite{hochreiter1997long}.
Similar to LSTM, TAMKOT is composed of a memory cell, an input gate, an output gate, and a forget gate. 
%At time step $t$, the hidden state from the previous step $t-1$ is used for student knowledge transfer and update. % each gate or cell takes the hidden state from $t-1$ as the input to transfer the knowledge. 
However, unlike LSTM which is invariant to activity types, TAMKOT models various activity types and their transitions by considering the current and previous activity type as an extra input, and adopting the internal gate formulations to appropriate activity type transitions. 
This results in a different formulation for each of TAMKOT's gate compared to LSTM and provides an explicit between-type knowledge transfer model.

In particular, we assume a different knowledge transfer pattern for each transition between learning material types. 
%Each of them represents the material types student interact with in two consecutive interactions.
%To illustrate, the knowledge that is transferred from watching video lectures to solve the problems could be different from the knowledge that is transferred from solving problems to understand the video lectures.
%For example, watching the lecture may help the student learn the knowledge to solve the problem immediately. But attempting the problem may not help the student understand the lecture. %, if the lecture has totally new concepts. 
For each of these transitions, we propose a set of indicators and 
%represent the learning material type permutations in the student sequence and 
formulate the gate updates according to these indicators. %update the  for each learning material permutation, we propose to utilize different weights for each gate or cell to transfer the knowledge.
For example, assuming video lectures (denoted by ``L'') and problems (denoted by ``Q'') as two material types, we have four different transitions between problems and video lectures: problems to problems (QQ), problems to video lectures (QL), video lectures to problems (LQ), and video lectures to video lectures (LL).
Consequently, we denote the four permutation indication variables at each time step $t$ according to the learning material type indicators $d_t$ and $d_{t-1}$:
%\textcolor{blue}{Consequently, at each time step $t$, we propose four indication variables represent the transition permutations according to the learning material type indicators $d_t$ and $d_{t-1}$:}

\vspace{-10pt}
{\footnotesize 
\begin{align}
\vspace{-10pt}
    &s_{QQ} = (1-d_t)(1-d_{t-1})\\
    &s_{QL} = d_t(1-d_{t-1})\\
    &s_{LQ} = (1-d_t)d_{t-1}\\
    &s_{LL} = d_td_{t-1} % for higher dimensions: d_t: 1-hot rep of type, S: matrix of outer product of d_t and d_t-1
    \vspace{-10pt}
\end{align}
}
%at each time step $t$, from learning material type indicator $d_t$ and $d_{t-1}$, TAMKOT knows the learning material type of $t$ and the learning material type of $t-1$. 
%Suppose there are two learning material types question and lecture, we could get four permutations. that can later indicate the correct weights to be used in each gate or cell:
where $s_{QQ}$, $s_{QL}$, $s_{LQ}$, and $s_{LL}  \in \{0, 1\}$ indicate four transition permutations of learning material type from time step $t-1$ to $t$.
%If, $s_{**} = 1$ and $s_** = 0$ represent whether or not the corresponding transition permutation is activated.
For example, $s_{QL}=1$ indicates the student has switched from attempting a problem at time $t-1$ to watching a video lecture at time $t$. %, and $s_{QQ}=1$ shows that the student attempts solving problems at both steps $t-1$ and $t$.
%both learning material type is questions, from learning material type questions to lecture, from learning material type lectures to questions, and lectures for both learning material type.
%Recall that $d_t = 0$ represents the learning material type of time step $t$ is question, and $d_t = 1$ for interaction of lecture learning material type.
As a result of this formulation, at each time step $t$, only one of four permutation indication variables is equal to $1$, with the rest of them being $0$. 
%For example, while the learning material type of $t$ and $t - 1$ are both question, $s_{QQ} = (1-0)(1-0) = 1$ and $s_{QL} = s_{LQ} = s_{LL} = 0$. 
% To model the knowledge transfer associated with each transition, we consider a separate transition weight matrix that is activated by the related indicator variable.

We use a vector $\mathbold{h}_{t} \in \mathbb{R}^{d_h}$ to keep track of student knowledge state at time step $t$, where $d_h$ is the hidden dimension size.
At each step, we update $\mathbold{h}_{t}$ according to the previous state $\mathbold{h}_{t-1}$ and the embedding vector of activity that the student has attempted ($\mathbold{x}_t$ and $\mathbold{l}_t$). 
%  this vector as the students try different learning activities. To represent knowledge transfer 
Additionally, since we assume that the transition order between activity types are important in how students learn, we update $\mathbold{h}_{t}$ according to the transition permutations $s_{**}$ defined above.
To represent how the knowledge transfers between activity types, we use transition-specific weight matrices (indicated by $\mathbold{W}$s) to update the student state.
% Accordingly, we formulate gates and memory cell for TAMKOT as follows: 
Accordingly, at each time step $t$, TAMKOT updates $\mathbold{h}_{t}$ by %uses the embedding vector of activity ($\mathbold{x}_t$ and $\mathbold{l}_t$) and the student hidden knowledge state $\mathbold{h}_{t-1}$ at last time step $t-1$ as the input and 
%formulating gates and memory cells 
as follows:

{\footnotesize		 
\begin{align}
\vspace{-5pt}
\begin{split}
    \mathbold{i}_t =& \sigma\Big((1 - d_t)\cdot\mathbold{x}_t\mathbold{V}_{iQ} + d_t \cdot \mathbold{l}_t \mathbold{V}_{iL} + s_{QQ}\cdot \mathbold{h}_{t-1}\mathbold{W}_{i_{QQ}}+ \\
    &s_{LL} \cdot \mathbold{h}_{t-1} \mathbold{W}_{i_{LL}} + s_{QL} \cdot \mathbold{h}_{t-1}\mathbold{W}_{i_{QL}} + s_{LQ}\cdot \mathbold{h}_{t-1}\mathbold{W}_{i_{LQ}} + \mathbold{b}_i\Big) \label{eq:inputgate}
\end{split}\\
\begin{split}
    \mathbold{g}_t =& tanh\Big((1 - d_t)\cdot\mathbold{x}_t\mathbold{V}_{gQ} + d_t \cdot \mathbold{l}_t \mathbold{V}_{gL} + s_{QQ}\cdot \mathbold{h}_{t-1}\mathbold{W}_{g_{QQ}} + \\
    &s_{LL} \cdot \mathbold{h}_{t-1} \mathbold{W}_{g_{LL}} + s_{QL} \cdot \mathbold{h}_{t-1}\mathbold{W}_{g_{QL}} + s_{LQ}\cdot \mathbold{h}_{t-1}\mathbold{W}_{g_{LQ}} + \mathbold{b}_g\Big)\label{eq:memorycell}
\end{split}\\
% \vspace{-5pt}
% \end{align}
% \begin{align}
% \vspace{-5pt}
% \begin{split}
%     \mathbold{i}_t =& \sigma\Big((1 - d_t)\cdot\mathbold{x}_t\mathbold{V}_{iQ} + d_t \cdot \mathbold{l}_t \mathbold{V}_{iL} + s_{QQ}\cdot \mathbold{h}_{t-1}\mathbold{W}_{i_{QQ}}+ \\
%     &s_{LL} \cdot \mathbold{h}_{t-1} \mathbold{W}_{i_{LL}} + s_{QL} \cdot \mathbold{h}_{t-1}\mathbold{W}_{i_{QL}} + s_{LQ}\cdot \mathbold{h}_{t-1}\mathbold{W}_{i_{LQ}} + \mathbold{b}_i\Big) \label{eq:inputgate}
% \end{split}\\
% \begin{split}
%     \mathbold{g}_t =& tanh\Big((1 - d_t)\cdot\mathbold{x}_t\mathbold{V}_{gQ} + d_t \cdot \mathbold{l}_t \mathbold{V}_{gL} + s_{QQ}\cdot \mathbold{h}_{t-1}\mathbold{W}_{g_{QQ}} + \\
%     &s_{LL} \cdot \mathbold{h}_{t-1} \mathbold{W}_{g_{LL}} + s_{QL} \cdot \mathbold{h}_{t-1}\mathbold{W}_{g_{QL}} + s_{LQ}\cdot \mathbold{h}_{t-1}\mathbold{W}_{g_{LQ}} + \mathbold{b}_g\Big)
% \end{split}\\
\begin{split}
    \mathbold{f}_t =& \sigma\Big((1 - d_t)\cdot\mathbold{x}_t\mathbold{V}_{fQ} + d_t \cdot \mathbold{l}_t \mathbold{W}_{fL} + s_{QQ}\cdot \mathbold{h}_{t-1}\mathbold{W}_{f_{QQ}} + \\
    &s_{LL} \cdot \mathbold{h}_{t-1} \mathbold{W}_{f_{LL}} + s_{QL} \cdot \mathbold{h}_{t-1}\mathbold{W}_{f_{QL}} + s_{LQ}\cdot \mathbold{h}_{t-1}\mathbold{W}_{f_{LQ}} + \mathbold{b}_f\Big)\label{eq:forgetgate}
\end{split}\\
\begin{split}
    \mathbold{o}_t =& \sigma\Big((1 - d_t)\cdot\mathbold{x}_t\mathbold{V}_{oQ} + d_t \cdot \mathbold{l}_t \mathbold{V}_{oL} + s_{QQ}\cdot \mathbold{h}_{t-1}\mathbold{W}_{o_{QQ}} + \\
    &s_{LL} \cdot \mathbold{h}_{t-1} \mathbold{W}_{o_{LL}} + s_{QL} \cdot \mathbold{h}_{t-1}\mathbold{W}_{o_{QL}} + s_{LQ}\cdot \mathbold{h}_{t-1}\mathbold{W}_{o_{LQ}} + \mathbold{b}_o\Big)\label{eq:outputgate}
\end{split}\\
  \mathbold{m}_t &= \mathbold{f}_t \cdot \mathbold{m}_{t-1} + \mathbold{i}_t \cdot \mathbold{g}_t\\ \label{eq:candidatememorycell}
  \mathbold{h}_t &= \mathbold{o}_t \cdot tanh(\mathbold{m}_t) 
\end{align}
}
where $\mathbold{i}_t$, $\mathbold{f}_t$, $\mathbold{o}_t$ represent the input gate, forget gate, and output gate, and $\mathbold{g}_t$ is the candidate memory cell, and $\sigma$ is the sigmoid function.
In Equations~\ref{eq:inputgate} to~\ref{eq:outputgate}, gates are calculated according to the learning material type transitions in the student sequence.
% In Equations~\ref{eq:inputgate} to~\ref{eq:outputgate}, gates are activated according to the learning material type transitions in the student sequence and their associated weights.  
Since knowledge transfer can be different for the four possible transitions, we consider separate transfer weight matrices for them.
So, in each gate and cell, $\mathbold{W}_{*_{QQ}}$, $\mathbold{W}_{*_{LL}}$, $\mathbold{W}_{*_{QL}}$, and $\mathbold{W}_{*_{LQ}}\in \mathbb{R}^{d_h\times d_h}$ are knowledge transfer weight matrices that are associated with the four different possible transition permutations $s_{**}$.
For example, $\mathbold{W}_{f_{LQ}}$ captures the knowledge transfer from previous student knowledge state ${h}_{t-1}$ to the current state in the forget gate, when a student switches from watching video lectures to solving problems. 
Also, $\mathbold{V}_{*Q} \in \mathbb{R}^{(d_q+d_a)\times d_h}$ and $\mathbold{V}_{*L}  \in \mathbb{R}^{d_l\times d_h}$ are used to map embeddings of problem activity and lecture activity, respectively, to gates and cell in the hidden knowledge layer.
$\mathbold{V}_{*Q}$ and $\mathbold{V}_{*L}$ are activated according to the current learning material type ($d_t$).
$\mathbold{b}_*\in \mathbb{R}^{d_h}$ are the bias terms.

% one activity 
% We represent knowledge transfer between different activity types
% \textcolor{blue}{To model the knowledge transfer associated with each transition, 
% we assume the amount of knowledge transfer from knowledge state $\mathbold{h}_{t-1}$ to $\mathbold{h}_{t}$ is different for different transition permutation of material types,
% and we consider separate transfer weight matrices that are activated by the related indicator variables $s_{**}$.
% If, $s_{**} = 1$ and $s_** = 0$ represent whether or not the corresponding transition permutation is activated.}

% \textcolor{blue}{In Equations~\ref{eq:inputgate} to~\ref{eq:outputgate}, gates are calculated according to the learning material type transitions in the student sequence.}
% , \textcolor{blue}{where $d_h$ is the hidden dimension size.}

% Similarly, the weights $\mathbold{V}_{*Q}$ and $\mathbold{V}_{*L}$ are employed according to the current learning material type ($d_t$).

Based on our formulation in TAMKOT, at each time step, one learning activity type transition, and consequently one knowledge transfer weight is activated.
%For example, if the student is working on a problem at time step $t$ and a video lecture at $t-1$, the input gate formulation in Equation~\ref{eq:inputgate} would be equivalent to $ \mathbold{i}_t = \sigma\Big(\mathbold{x}_t\mathbold{V}_{iQ} + \mathbold{h}_{t-1}\mathbold{W}_{i_{LQ}} + \mathbold{b}_i\Big)$. %TAMKOT counts the first term of equation \ref{eq:inputgate} for question embedding while the second term for lecture embedding is ignored, because of the $d_t = 0$ and $d_{t-1} = 1$ for this situation. 
%Then the permutation indication variables make all other terms from $\mathbold{h}_{t-1}$ to be 0. 
While we consider two learning material types in this paper, extending this formulation to more than two types would be trivial.
%To this end, TAMKOT can model the knowledge gain and transfer between different permutations and unlimited switches of different learning material types in the student learning sequence.
Unlike previous attempts at multi-activity knowledge tracing that allowed for a limited number of non-assessed learning materials between every two assessed ones~\cite{wang2021learning}, this representation allows us to model unlimited transitions in any order between two learning material types in student sequence.
Additionally, unlike sequential multitask learning models that need the same sequence length for all views~\cite{chen2018meta,liu2016context}, this representation does not need one-to-one sequence alignment and allows for different sequence lengths for assessed and non-assessed learning activities.

\subsection{Prediction Layer}
%Many previous knowledge tracing methods propose to predict a vector at each time step. The predicted vector represents the prediction of student's possible performance on each of the question.  
%However, in the real learning process, the student usually knows the question before they provide the answer. Accordingly, TAMKOT predicts one possible performance for the given question at each time step.
In this layer, TAMKOT predicts the target student's performance for a given problem $q_{t+1}$ at the next time step $t+1$, according to the student's past learning activities, summarized in the current hidden state $\mathbold{h}_{t}$.
%The hidden state $\mathbold{h}_{t}$ can be considered as the student's knowledge state at time step $t$.
%Through all the modeling of the four gates in the hidden knowledge transfer layer, TAMKOT gets the student's knowledge state $\mathbold{h}_t$ after the interaction at time step $t$. 
%We now want to use the student's current knowledge state $\mathbold{h}_t$ to predict the student performance on the next question $q_{t+1}$. 
This is achieved by concatenating the hidden state $\mathbold{h}_t$ with the next candidate problem's embedding vector $\mathbold{q}_{t+1}$, and passing the concatenation into a fully connected layer with the sigmoid activation function:

{\footnotesize 
\begin{equation}
    % \vspace{-5pt}
    p_{t+1} = \sigma(\mathbold{W}_{p}^T[\mathbold{h_t} \oplus \mathbold{q_{t+1}}] + b_p) \label{eq: prediction}
    %   \vspace{-5pt}
\end{equation}
}
where the prediction $p_{t+1}$ represents the probability that the student answers the next problem $q_{t+1}$ correctly, $\mathbold{W}_p \in \mathbb{R}^{d_q+d_h}$ is the weight matrix, and $b_p \in \mathbb{R}$ is the bias term. %, $\mathbold{W}_p \in \mathbb{R}^{d_q+d_h}$ and $b_p \in \mathbb{R}$ is the weight matrix and the bias term of the full connected layer. The $p_{t+1}$ is formulated into range $[0, 1]$ by the sigmoid activation function.

\subsection{Objective Function}
We learn the parameters of TAMKOT by  
%We joint learn all learnable parameters by 
minimizing the following regularized binary cross-entropy loss.

{\footnotesize 
\begin{equation}
   % \vspace{-5pt}
\mathcal{L}= -\sum_{t}\left(r_{t} \log p_{t}+\left(1-r_{t}\right) \log \left(1-p_{t}\right)\right) + \lambda_{\theta}||\theta||^2 \label{eq:objectivefunction}
\end{equation}
%   \vspace{-5pt}
}
 %over all training students' performances $r_t$ and prediction $p_{t}$. 
where $r_t$ represents the actual student performance and $\theta$ denotes all the learnable parameters of TAMKOT: the embedding matrices $\mathbold{A}^q$, $\mathbold{A}^l$, and $\mathbold{A}^r$, the weight matrices $\mathbold{V}_{*Q}$, $\mathbold{V}_{*L}$, $\mathbold{W}_{*_{QQ}}$, $\mathbold{W}_{*_{LL}}$, $\mathbold{W}_{*_{QL}}$, and $\mathbold{W}_{*_{LQ}}$, and the bias terms $\mathbold{b}_*$. 
$||\theta||^2$ is the regularization term and $\lambda_\theta$ is the hyperparameter to specify regularization weight. 
%It is also worth noting, TAMKOT evaluates only the time steps of the assessed learning material type.  
%For those time steps $t$ that the learning material type is non-assessed, we cannot have the student's actual performance, TAMKOT masks and ignores those time steps when calculating the objective function.

% \begin{figure*}[!ht]
    % \centering
    % \subfigure{
    % \includegraphics[width=0.99\textwidth]{figures/model2.pdf}}
    % \caption{model}
    % \label{fig:model}
% \end{figure*}

\section{Experiments}
\label{sec:experiments}
% We evaluate TAMKOT with four kinds of experiments. 
% First, we evaluate TAMKOT in the student performance prediction task compared to fourteen state-of-art KT baselines.
% Second, to analyze the knowledge transfer between assessed and non-assessed learning material types, we visualize and compare the transition matrices for different permutations of learning material types. 
% Third, to check if the model represents reasonable student knowledge states, we perform a case study to visualize student knowledge states at each time step. 
% Finally, to validate if the model discovers meaningful latent concepts in both assessed and non-assessed learning materials types, we run a clustering analysis of discovered embedding matrices. Our code and data will be released publicly upon acceptance.
We evaluate TAMKOT with three sets of experiments. 
First, we evaluate TAMKOT in the student performance prediction task.
Second, to analyze the knowledge transfer between assessed and non-assessed learning material types, we compare the transition matrices for different transitions.
Third, we perform a case study to visualize student knowledge states. % to check if the model represents reasonable student knowledge states.
%Finally, we run a clustering analysis of discovered embedding matrices to validate if the model discovers meaningful latent concepts of learning materials. 
Our code and example datasets are available on GitHub~\footnote{\url{https://github.com/persai-lab/BigData2022-TAMKOT}}.

\begin{table*}[!htbp]
\centering
% \vspace{-5pt}
\caption{Descriptive statistics of 3 datasets.}
\vspace{-5pt}
\resizebox{0.78\textwidth}{!}{
\begin{tabular}{cccccccccc}
\toprule
Dataset &\#Users &\#Questions &\begin{tabular}[c]{@{}c@{}} Question\\Records\end{tabular} &\begin{tabular}[c]{@{}c@{}} Question\\Responses\\Mean\end{tabular} &\begin{tabular}[c]{@{}c@{}} Question\\Responses\\STD\end{tabular} &\begin{tabular}[c]{@{}c@{}} \#Correct\\Question\\Responses\end{tabular} &\begin{tabular}[c]{@{}c@{}} \#Incorrect\\Question\\Responses\end{tabular} &\begin{tabular}[c]{@{}c@{}} \#Non-assessed\\materials\end{tabular}& \begin{tabular}[c]{@{}c@{}} \#Non-assessed\\ Records\end{tabular} \\
\hline
MORF &$686$ &$10$ &$12031$ &0.7763 &0.2507 &N/A &N/A &$52$ &$41980$\\ 
EdNet &$1000$ &$11249$ &$200931$ &$0.5910$ &$0.2417$ &$118747$ &$82184$ &$8324$ &$150821$\\ 
% Junyi1564 &$1564$ &$142$ &$120984$ &N/A &N/A &$86654$ &$34328$ &$116$ &$16389$ \\ 
Junyi &$2063$ &$3760$ &$290754$ &$0.6660$ &$0.2224$ &$193664$ &$97090$ &$1432$ &$69050$ \\ 
\bottomrule
\end{tabular}
}
\label{table:datasets}
\vspace{-5pt}
\end{table*}

\subsection{Datasets}
We use three real-world datasets in our study. 
The general statistics of each dataset can be found in
Table \ref{table:datasets}.\\ 
\textbf{MORF}\footnote{\url{https://educational-technology-collective.github.io/morf/}}~\cite{andres2016replicating}: This is a dataset of one online course, available via the MOOC Replication Framework (MORF), from Coursera\footnote{\url{https://www.coursera.org/}}.
The course subject is `educational data mining' and is divided into different modules.
Each module is associated with a topic, such as `classification'.
% We use two learning material types from this course: video lectures (non-assessed), and assignments (assessed). 
We use video lectures (non-assessed), and assignments (assessed) as two learning material types.
In each module that is planned for a week, the students need to watch five to seven video lectures and work on one assignment.
Each assignment usually contains more than one problem. 
Only coarse-grained assignment-level data is available, as if the
students submit an entire assignment each time rather than submitting a single problem. 
We treat each student submission of assignment as one assessed activity and the overall score of each submission as the response to this activity. \looseness=-1
% We consider student every submission and the total score of entire assignment as one learning activities.
% Only student's total score of assignments are available.
%The students also can communicate with the instructor and their classmates in the discussion forum. 
%In this paper, we focus on the students' interactions of submitting assignments (assessed) and watching the video lectures (non-assessed).

\textbf{EdNet}\footnote{\url{https://github.com/riiid/ednet}}~\cite{choi2020ednet}: This is a dataset from a multi-platform AI tutoring service (Santa~\footnote{\url{https://www.aitutorsanta.com/}}) for Korean students to practice while preparing for TOEIC~\footnote{\url{https://www.ets.org/toeic}} English testing. 
% Santa collected all student-system interactions over 2 years. 
% It includes all student-system interactions collected over 2 years.
Ednet offers four different levels of data to provide various kinds of actions in a consistent and organized manner.
Data from the third level is selected to evaluate our model, which consists of student learning activities in multiple learning material types.
% Santa provides various learning material types, including questions (assessed), and their associated problem explanations (non-assessed).
%Students can choose the topic they want to practice. 
% We randomly sample $1000$ students who interacted with two learning material types in this dataset: questions (assessed), and their associated problem explanations (non-assessed).
During the student practice, the platform recommends questions to students. 
But, the students can decide to follow the recommendations or not.
Each question has a problem explanation that the students can choose to read.
We randomly sample $1000$ students who interacted with both questions (assessed) and their associated problem explanations (non-assessed) in this dataset.
% During the student practice, the platform recommends questions to students. 
% But, the students can decide to follow the recommendations or not.
% Each question has a problem explanation that the students can choose to read.
%We consider the question as the assessed learning material type and problem explanation as the non-assessed learning material type for experiments.
% There are $297,915$ students in the full dataset, out of which we randomly select $1000$ sample students to evaluate the proposed model.

\textbf{Junyi}\footnote{\url{https://pslcdatashop.web.cmu.edu/DatasetInfo?datasetId=1275}}~\cite{Junyi}: This dataset is collected from a Chinese e-learning website to teach students math. 
Students work on studying eight math areas with different difficulty levels. They start from the easiest level and are moved to the more difficult levels as they learn. 
% We download the preprocessed data introduced in~\cite{chang2015modeling} and use two learning material types in this dataset: problems (assessed), and hints (non-assessed). 
% In this dataset, each problem may be associated with more than one hint. 
% The students can request hints for solving the problems.
We use the preprocessed data introduced in~\cite{chang2015modeling}. 
In this dataset, problems (assessed), and hints (non-assessed) are used as two learning material types.
During the practice, students have the option to request hints for solving the problems, and each problem may be associated with more than one hint.
%The student can choose to look through all the hints or guess all the possible answers.
%We utilize the student requests as the non-assessed learning material type. 
% There are a total of $247,606$ students in the dataset. We randomly select $2063$ students and use their learning sequences for our experiments. %$1/120$ portion of students as the dataset for our experiments.

\subsection{Baseline Methods}
\label{sec:baseline}
We utilize six state-of-art assessed-only supervised KT models and two multi-activity KT models as original baselines to evaluate our proposed method.
To provide a fair comparison, we also extend the six assessed-only supervised KT models to be able to consider both assessed and non-assessed learning material types and used them as our baselines.
%Additionally, we developed multi-type variants of the six supervised KT models to use as out baseline. 
% We compare our proposed method with 14 baselines based on nine original state-of-art KT models.
In addition, we also extend the simple multi-layer (MLP) perceptron as another baseline that can incorporate both assessed and non-assessed learning activities as the input.
These baselines are identified by a ``+M'' at the end of their names.
% Among all original baselines, only the two multi-type ones explicitly model learning activities of both assessed and non-assessed learning material types.
% The other 6 are designed only based on the assessed learning material types.
% To provide a fair comparison, we extend the assessed-only KT models to be able to consider both assessed and non-assessed learning material types.
In total, we compare our TAMKOT with 14 baselines.
Seven of the eight original baselines are based on deep learning, one is a tensor factorization model.
The assessed supervised KT baselines are:\\
$\bullet$DKT~\cite{piech2015deep} is the first deep learning based KT model that uses recurrent neural networks to model student knowledge gain. \\
$\bullet$DKVMN~\cite{zhang2017dynamic} uses memory augmented neural networks to model KT, with one static key matrix for the knowledge concepts and a dynamic value matrix for updating student mastery levels.\\
$\bullet$DeepIRT~\cite{yeung2019deep} is an extension of DKVMN that integrates the one-parameter logistic item response theory to address overfitting and provide better interpretation.\\
$\bullet$SAKT~\cite{pandey2019self} applies the self-attentive model for KT to model the relationship between interactions at different time steps.\\
$\bullet$ SAINT~\cite{choi2020towards} is the first encoder-decoder model for KT that applies deep self-attentive layers to exercises and responses separately.\\
$\bullet$AKT~\cite{ghosh2020context} is a context-aware model that uses a monotonic attention mechanism to summarize past student performances that are relevant to the current problem.
% \begin{itemize}
%     \item DKT~\cite{piech2015deep} is the first deep learning based KT model that uses recurrent neural networks modeling student knowledge gain. 
%     \item DKVMN~\cite{zhang2017dynamic} uses memory augmented neural networks to model KT, with one static key matrix for the knowledge concepts and a dynamic value matrix for updating student mastery levels.
%     \item DeepIRT~\cite{yeung2019deep} is a extension of DKVMN that integrates the one-parameter logistic item response theory (1PL-IRT) to address overfitting and provide better interpretation.
%     \item SAKT~\cite{pandey2019self} applies the self-attentive model for KT to model the relationship between interactions at different time steps.
%     \item SAINT~\cite{choi2020towards} applies deep self-attentive layers to exercises and responses separately. It is the first work to suggest an encoder-decoder model for KT.
%     \item AKT~\cite{ghosh2020context} is a context-aware model that uses a monotonic attention mechanism to summarize past student performances that are relevant to the current problem.
% \end{itemize}

In addition to the above, we compare TAMKOT with the following models that support both assessed and non-assessed learning material types:\\
$\bullet$MVKM~\cite{zhao2020modeling} is a multi-view tensor factorization method that explicitly models student knowledge acquisition from multi-type learning activities. It builds separate tensors for students' activities from each learning material type, and cannot explicitly model knowledge transition between material types.\\
$\bullet$DMKT~\cite{wang2021learning} also explicitly models students' knowledge gain from both assessed and non-assessed activities. It is based on DKVMN and models different read and write operations for assessed and non-assessed learning material types. However, it does not explicitly model knowledge transfer between assessed and non-assessed learning materials. Additionally, it only allows for a fixed number of non-assessed learning activities between every two assessed ones. As a result, it is not flexible to capture the full student sequence and switches between learning material types.\\
$\bullet$MLP+M is a simple multi-layer perceptron that considers a student's three recent assessed interactions along with three non-assessed interactions as the input and predicts the probability of student mastery level.\\
% $\bullet$DKT+M~\cite{zhang2017incorporating} and DKVMN+M are extensions of DKT and DKVMN to consider non-assessed learning activities in addition to the assessed ones. They summarize all the non-assessed learning materials that the student had interacted with between every two assessed activities, as an additional feature. The non-assessed learning materials' embeddings are concatenated with the problem embedding as input for vanilla DKT and DKVMN.\\
$\bullet$DKT+M~\cite{zhang2017incorporating} and DKVMN+M are extensions of DKT and DKVMN to consider non-assessed learning activities in addition to the assessed ones. 
They concatenate embedding vectors of all non-assessed learning materials that the student had interacted with between every two assessed activities as an additional feature, with the problem embedding as input for vanilla DKT and DKVMN. \\
$\bullet$SAINT+M~\cite{choi2020towards}, SAKT+M, and AKT+M are variants of SAINT, SAKT, and AKT. Similar to DKT+M, in these extended models, all the non-assessed learning materials' embeddings that happen between two assessed activities are summarized as an additional feature. In addition, the position encoding is added to each learning material embedding.\\

%Moreover, despite that some approaches originally proposed to use materials' knowledge concepts as input, we used learning materials as input for all of the above baselines for a fair comparison.
Notably, DKT+M is the closest approach to an ablated version of TAMKOT which does not include the knowledge transfer component %among different learning activities 
and ignores the knowledge transition between non-assessed learning activities. 

% DKT+M is the closest approach to TAMKOT, which does not include knowledge transfer among different learning activities and ignores the knowledge transition between non-assessed learning activities. 
% For fair comparison, we use assessed material and response or lectures as the input for all methods
\begin{table}[!htbp]
\centering
\vspace{-10pt}
\caption{Learned Best Hyperparameters of TAMKOT}
\vspace{-5pt}
\resizebox{0.3\textwidth}{!}{
\begin{tabular}{ccccccc}
\toprule
Dataset &$d_q$ & $d_l$ & $d_a$ & $d_h$ & $\mathnormal{L_s}$ & $\lambda_{\theta}$ \\
\hline
MORF &64 & 16 & 64 & 16& 100 &0.01 \\ 
EdNet &64 & 32  &64  &16 &50  &0.01 \\ 
% Junyi1564 & &  &  & &  & \\ 
Junyi &32 & 32 &32  &32 &100  &0.05 \\ 
\bottomrule
\end{tabular}
}
\label{table:hyperparameters}
\vspace{-12pt}
\end{table}
% The paper is interesting and well-written. However, there are a couple of important factors needed for clarification.
% 1) It is not clear to me how the authors compare with the models (such as DKT and DKVMN), that take knowledge components as their inputs. Do they also take (questions and responses) or lectures as input?
% explain in one sentence

\subsection{Experiment Setup}
We use 5-fold student stratified cross-validation to split the training set and test set. 
At each fold, sequences from $80\%$ of the students are used as the training set, and the sequences from the rest $20\%$ of students are used as the testing set. 
For hyperparameter tuning, we separate another $20\%$ of students from the training set as the validation set. 
We implement our proposed methods with PyTorch~\footnote{https://pytorch.org/} and use the Adam optimizer to learn the model parameters. 
All parameters are randomly initialized with the Gaussian distribution with $0$ mean and $0.2$ standard deviation.
We use the norm clipping threshold to avoid gradient exploding. 
Following the standard KT experiments~\cite{piech2015deep}, we truncate or pad all the sequences to the same length. 
Sequence length ($\mathnormal{L_s}$) is treated as another hyperparameter fine-tuned using the validation data.
For sequences longer than $\mathnormal{L_s}$, we truncate it into multiple sequences.
For sequences shorter than $\mathnormal{L_s}$, we pad them to length $\mathnormal{L_s}$ with $0$. 
% To find the best hyperparameters, reported in table \ref{table:hyperparameters}, we use coarse-grained grid search. 
We use coarse-grained grid search to find the best hyperparameters (reported in table \ref{table:hyperparameters}).

\subsection{Student Performance Prediction}
\label{sec:experiments:performance}

Here, we evaluate TAMKOT on the task of student performance prediction with the baselines introduced in Section~\ref{sec:baseline}.
We report average results across the five folds, as well as t-test p-values compared with the proposed model TAMKOT. 
In EdNet and Junyi datasets, student responses are binary (success or failure).
So, we use Area Under Curve (AUC) to evaluate model performances.
Higher AUC represents better prediction performance. 
In the MORF dataset, assignments are graded using a numeric value. 
% We normalize students' assignment scores in $[0,1]$ with the maximum possible score for the assignment, and use the normalized scores as student performance.
We normalize students' assignment scores in range $[0,1]$ with the maximum possible score for the assignment as student performance.
Root Mean Squared Error (RMSE) is used to evaluate the prediction performance of the MORF dataset.
Lower RMSE accounts for better prediction performance. 
Experiments results are presented in table~\ref{table:results}. 
Since MVKM is lacking in handling high-dimensional datasets with a high computation time, we only run MVKM on the MORF dataset.
\begin{table}[!htbp]
\vspace{-5pt}
% \caption{Student Performance Prediction Results. We report RMSE for MORF dataset and AUC for EdNet and Junyi datasets. The best and second-best result are in boldface and underline, respectively. $**$ and $*$ indicate paired t-test p-value<0.05 and p-value<0.1, respectively, compared to TAMKOT.}
\caption{Student Performance Prediction Results. The best and second-best result are in boldface and underline, respectively. $**$ and $*$ indicate paired t-test $p-value<0.05$ and $p-value<0.1$, respectively, compared to TAMKOT.}
\vspace{-5pt}
\centering \scriptsize
\resizebox{0.4\textwidth}{!}{
\begin{tabular}{c|c|c|c}
\toprule
& MORF & EdNet  & Junyi\\
\cline{2-4}
Methods & RMSE & AUC &AUC \\ 
\midrule
DKT &$0.1938^{*}$  &$0.6393^{**}$ &$0.8623^{**}$\\ 
DKVMN &$0.2043^{**}$ &$0.6296^{**} $ &$0.8558^{**}$\\ 
SAKT &$0.2113^{**}$ &$0.6334^{**} $ &$0.8053^{**}$\\
SAINT &$0.2019^{**}$ &$0.5205^{**}$ &$0.7951^{**}$\\
AKT & $0.2420^{**}$ &$0.6393^{**} $ &$0.8093^{**}$\\
DeepIRT &$0.1946^{**}$ &$0.6290^{**}$ &$0.8498^{**}$\\ 
\midrule
DKT+M & $0.1928$ &$0.6372^{**}$ &$\underline{0.8652^{*}}$\\
DKVMN+M & $0.2251^{**}$ &$0.6343^{**}$ &$0.8513^{**}$\\
SAKT+M & $0.2085^{**}$ &$0.6323^{**}$ &$0.7911^{**}$\\
SAINT+M & $0.1977^{**}$ &$0.5491^{**}$ &$0.7741^{**}$\\
AKT+M & $0.2240^{**}$ &$\underline{0.6404^{**}}$ &$0.8099^{**}$\\
MLP+M & $0.2433^{**}$ &$0.6102^{**}$ &$0.7290^{**}$\\
\midrule
MVKM &$0.1936^{*}$ &$-$ &$-$\\
DMKT &$\mathbf{0.1754^{**}}$& $0.6394^{**}$ &$0.8561^{**}$\\
\midrule
TAMKOT &$\underline{0.1871}$& $\mathbf{0.6786}$&$\mathbf{0.8745}$\\
\bottomrule
\end{tabular}
}
% \vspace{-5pt}
\label{table:results}
\end{table}

We see that TAMKOT significantly outperforms all the six supervised assessed KT models in all datasets.
This shows that \textit{TAMKOT can successfully model the non-assessed student activities along with the assessed ones, to leverage their added information for improving the performance predictions.}
The other two explicit multi-activity KT models MVKM, and DMKT mostly achieve higher prediction performance compared to the assessed-only methods.
This shows that explicitly modeling non-assessed learning activities could help improve KT.
%Particularly, DMKT performs better than the assessed-only methods in the MORF dataset. 
But, we also see that DKT and AKT perform better than or similar to DMKT in EdNet and Junyi datasets.

One potential reason for this observation could be the difference between MORF and the other two datasets.
The learning materials in the MORF dataset are more complex, compared to EdNet and Junyi.
While in EdNet and Junyi problems are granular and focused on specific topics, each MORF assignment includes multiple problems, each of which covers multiple concepts.
We note that DMKT's structure is also more complex than DKT and AKT.
While DKT and AKT use a vector representation $\mathbf{h}_t$ for student state, DMKT uses a complex key-value memory matrix representation for learning material and student knowledge.
Accordingly, we hypothesize that DMKT's better performance in MORF dataset can be attributed to a better match of DMKT's complexity with MORF's material complexity.
At the same time, this complexity may not be necessary for the EdNet and Junyi datasets with simpler learning material structures.

Comparing TAMKOT with the six multi-activity versions of the assessed KT models (the "+M" methods), we see that TAMKOT significantly outperforms all of them in all datasets.
Particularly, one can consider the LSTM-based DKT+M as a simpler version of TAMKOT without explicitly modeling transitions between activity types. 
TAMKOT significantly outperforms DKT+M in all datasets.
This shows that \textit{merely concatenating the assessed activity sequences with the non-assessed ones is not enough.}

In fact, we find that \textit{simply incorporating the non-assessed learning material as additional features can sometimes harm the prediction performance.} % does not always improve prediction performance. 
For example, the results of DKVMN+M are worse than DKVMN on all the datasets, DKT+M performs worse than DKT in the EdNet dataset, %only achieves a slightly better but not significant prediction performance on the MORF and Junyi datasets. 
and SAKT+M, SAINT+M, and AKT+M are worse than SAKT, SAINT, and AKT respectively in the Junyi dataset. 
%Moreover, SAKT+M, SAINT+M, and AKT+M also integrate the information from the interaction position.
%Although this helps improve prediction performance on the MORF and EdNet datasets, it hurts the prediction results in the Junyi dataset.
% Moreover, SAKT+M, SAINT+M, and AKT+M integrate the information from non-assessed learning materials as well as the interaction position. However, this hurts the prediction results in the Junyi dataset. 
%This shows that simply adding the non-assessed learning materials and their position as additional features is not helpful to accurately model student knowledge, and modeling the learning material transitions is necessary for it. 

Comparing TAMKOT with the two multi-activity baselines, we see that it outperforms both of them in EdNet and Junyi datasets. 
This shows that, \textit{modeling knowledge transfer and activity transitions, is essential in multi-activity knowledge modeling in these datasets.}
%TAMKOT's modeling of knowledge transfer and activity transition helps in student performance prediction in these datasets.
In the MORF dataset, TAMKOT %significantly outperforms all assessed KT models, their multi-type versions, and the multi-type MVKM, and 
is the second-best after DMKT.
%However, it is the second best in the MORF dataset, where DMKT performs better than TAMKOT.
%This shows that despite its simplicity, TAMKOT can perform relatively well in the MORF dataset too.
We hypothesize that this happens because of the MORF learning material complexity reason explained above. 
Similar to DKT and AKT, TAMKOT uses a simple vector-representation $\mathbf{h}_t$ for student state.
%One potential reason for the better performance of DMKT could be the difference between MORF and the other two datasets.
%The learning materials in the MORF dataset are more complex, compared to EdNet and Junyi.
%While in EdNet and Junyi problems are atomic and focused on specific topics, each MORF assignment includes multiple problems, each of which covers multiple concepts.
%Accordingly, we hypothesize that DMKT's better performance in MORF dataset can be attributed to a better match of DMKT's complexity with MORF's material complexity.
%In comparison with DMKT in the MORF dataset, w
%We note that unlike TAMKOT that uses a vector representation $\mathbf{h}_t$ for student state,
%DMKT uses a complex key-value memory matrix representation for learning material and student knowledge.

Overall, explicitly modeling both assessed and non-assessed activities, in addition to the transition-aware knowledge transfers between them, is shown to be necessary to accurately represent student knowledge and predict their performance.

% However, assessed-only DKT outperforms multi-type DMKT in the Junyi dataset.
% We also observe that most of the models achieve good prediction results (around $0.8$ AUC) in this dataset. 
% This could mean that the Junyi dataset is easy to model, potentially leading to overfitting in complex models, like DMKT and DKVMN.
%Complex methods such as DMKT and DKVMN may easily overfit the data that harm the predictions.
%The possible reason could be, the Junyi dataset is easy to model, most of baselines can get good prediction results (around 0.8 AUC). Complex methods such as DMKT and DKVMN may easily overfit the data that harm the predictions.

\subsection{Knowledge Transfer Analysis}
\label{sec:experiments:trainsion}
\begin{figure}[!ht]
    \centering
    % \subfigure[Weight matrices for MORF]{
    % \includegraphics[width=0.25\textwidth]{figures/model_analysis_forget_gate_weight_normalized_addtitle_V.pdf}\label{fig:trainsition_weight_morf}}
    % \subfigure[Weight matrices for EdNet]{
    % \includegraphics[width=0.25\textwidth]{figures/Ednet/model_analysis_forget_gate_weight_normalized_addtitle_V.pdf}\label{fig:trainsition_weight_ednet}}
    % \subfigure[Weight matrices for Junyi]{
    % \includegraphics[width=0.25\textwidth]{figures/Junyi/model_analysis_forget_gate_weight_normalized_addtitle_V.pdf}\label{fig:trainsition_weight_junyi}}
    % \caption{Heatmaps for comparison of the weight matrices $\mathbold{W}_{f_{QL}}$ (assessed to non-assessed type) and $\mathbold{W}_{f_{LQ}}$ (non-assessed to assessed type) of forget gate for each dataset}\\
    \subfigure[Forget gate weight matrices for MORF]{
    \includegraphics[width=0.8\columnwidth]{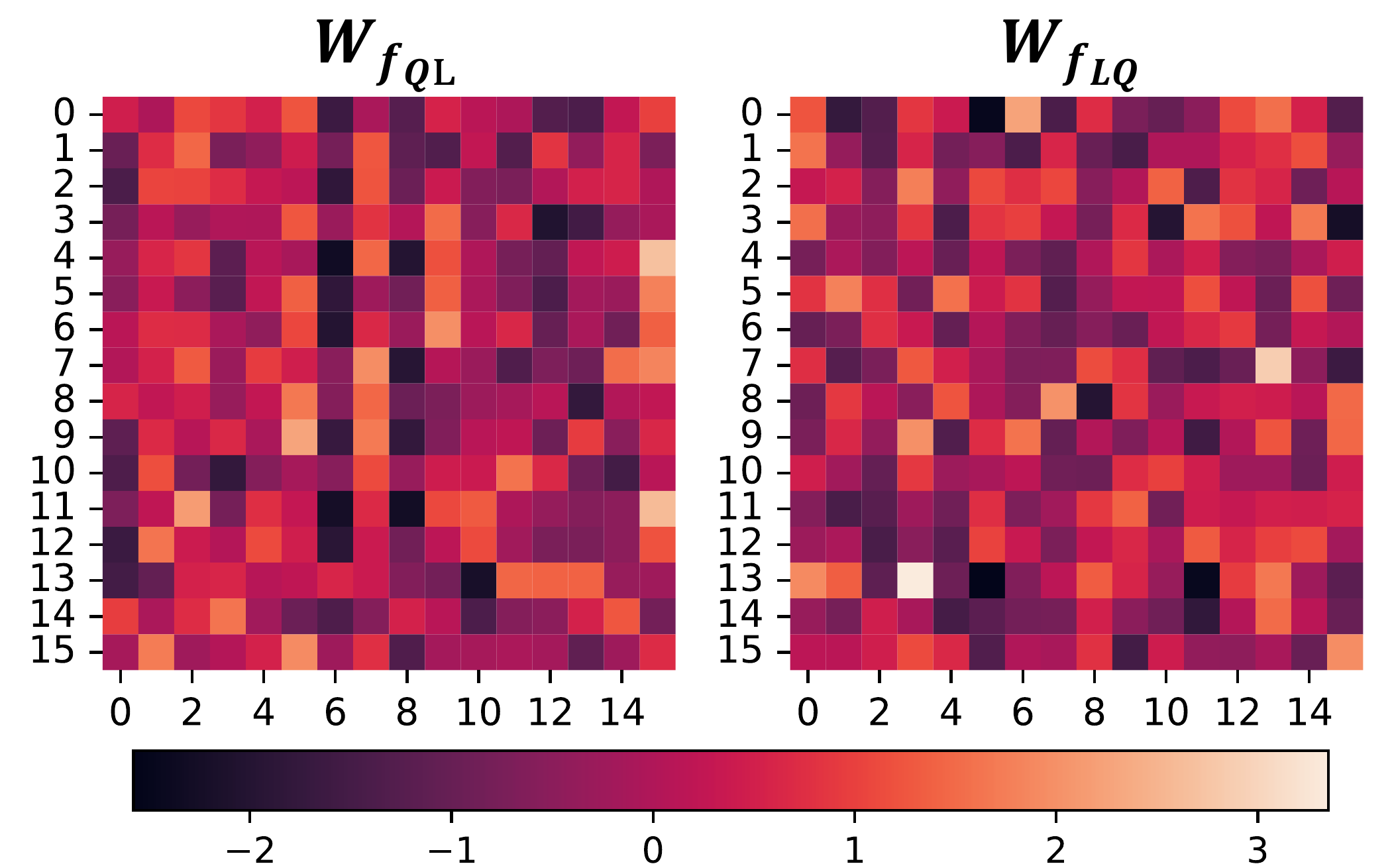}\label{fig:trainsition_weight_morf}}
    % \vspace{-10pt}
    \subfigure[Forget gate weight matrices for EdNet]{
    \includegraphics[width=0.8\columnwidth]{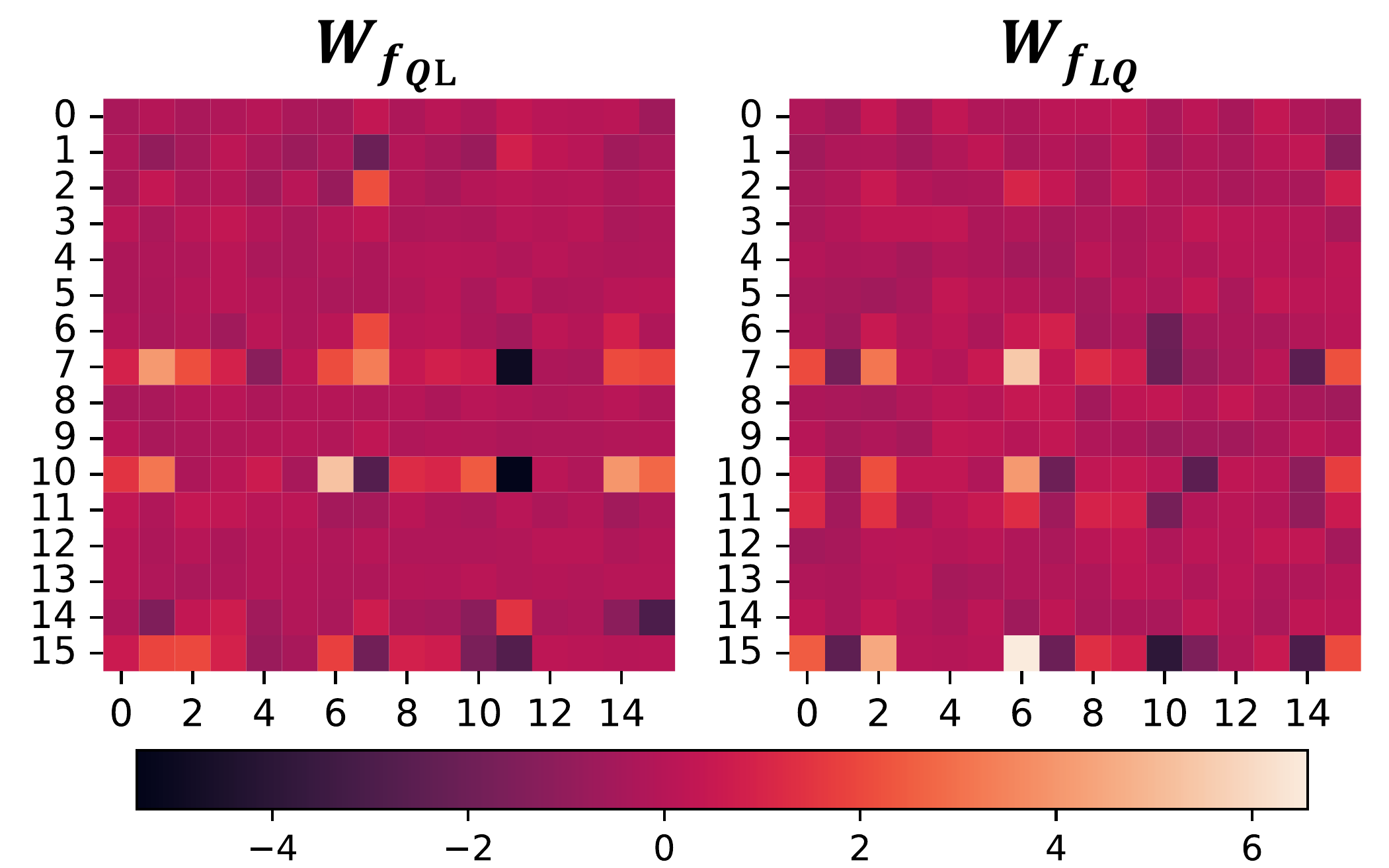}\label{fig:trainsition_weight_ednet}}
    % \subfigure[Weight matrices for EdNet]{
    % \includegraphics[width=0.33\textwidth]{figures/Ednet/EdNet.pdf}\label{fig:trainsition_weight_ednet}}
    % \vspace{-10pt}
    \subfigure[Forget gate weight matrices for Junyi]{
    \includegraphics[width=0.8\columnwidth]{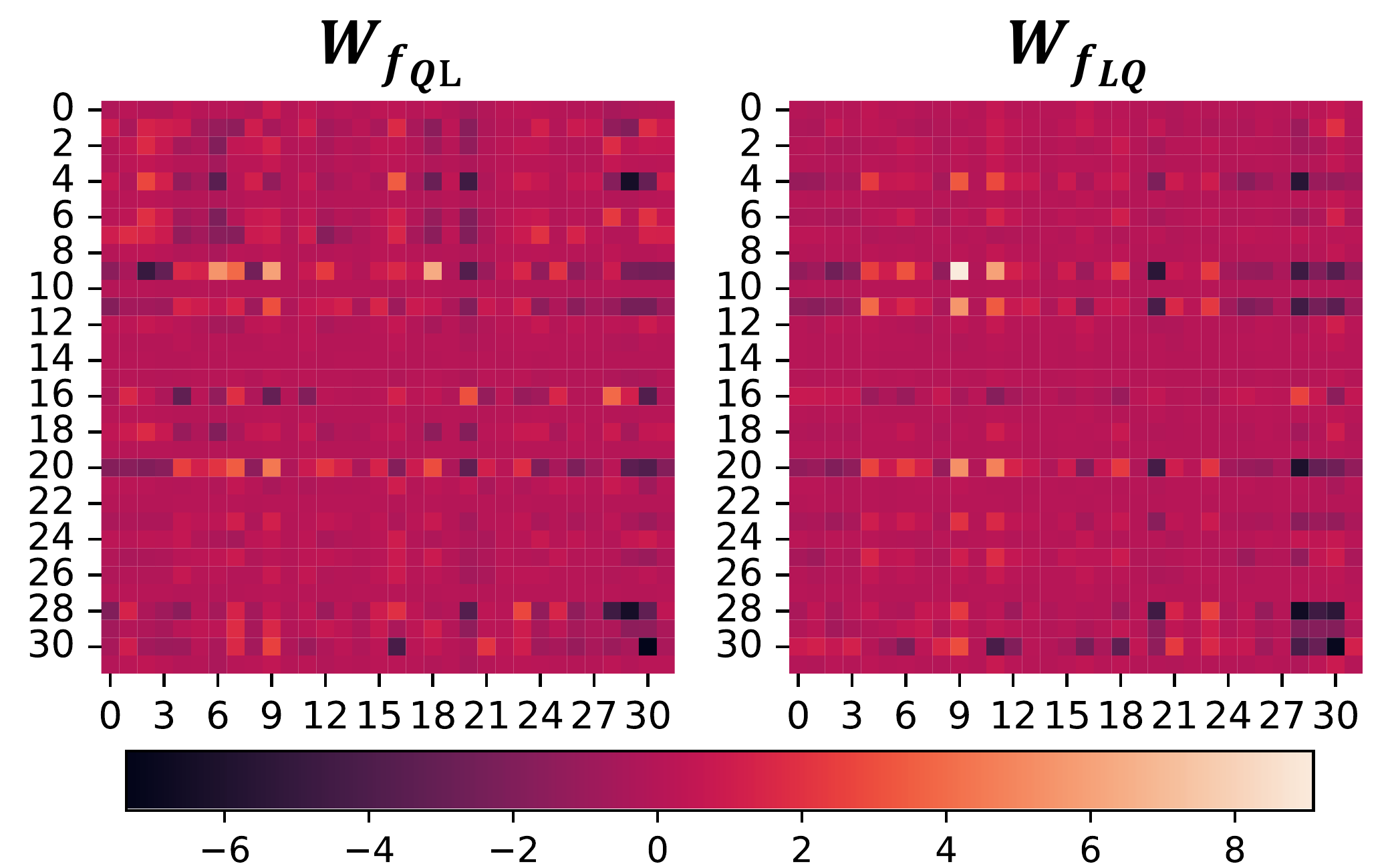}\label{fig:trainsition_weight_junyi}}
    \caption{Heatmaps for comparison of the weight matrices $\mathbold{W}_{f_{QL}}$ (assessed to non-assessed type) and $\mathbold{W}_{f_{LQ}}$ (non-assessed to assessed type) of forget gate for each dataset. 
    % $\mathbold{W}_{f_{QL}}$ (left) is the weight matrix of assessed to non-assessed type and $\mathbold{W}_{f_{LQ}}$ (right) is the weight matrix of non-assessed to assessed type.
    }
    \label{fig:transition_analysis}
    \vspace{-15pt}
\end{figure}

\begin{table}[!t]
\vspace{-5pt}
\centering
\caption{Spearman correlation coefficients with p-values between $\mathbold{W}_{f_{QL}}$ and $\mathbold{W}_{f_{LQ}}$}
\vspace{-5pt}
\resizebox{0.35\textwidth}{!}{
\begin{tabular}{c|ccc}
\toprule
            & MORF    & EdNet    & Junyi     \\
            \hline
Correlation & -0.03686  & 0.33680  & 0.38443 \\
p-value     & 0.55714 & 3.30e-08 & 2.09e-37 \\
\bottomrule
\end{tabular}} \label{fig:spearman}
\vspace{-10pt}
\end{table}

Here we analyze the learned knowledge transfer between assessed and non-assessed learning activity types. 
Particularly, we study if the knowledge transfer from the assessed learning materials to the non-assessed ones is different from the knowledge transfer from the non-assessed learning materials to the assessed ones.
We first inspect the transition weight matrices of the forget gate $\mathbold{W}_{f_{QL}}$ (assessed to non-assessed) and $\mathbold{W}_{f_{LQ}}$ (non-assessed to assessed) in equation~\ref{eq:forgetgate}.
%\textcolor{blue}{Since $\mathbold{W}_{f_{QL}}$ represents knowledge transfer weight from assessed materials to non-assessed materials in forget gate and $\mathbold{W}_{f_{LQ}}$ is from non-assessed materials to assessed materials.}
% First, we run a Wilcoxon signed-rank test~\cite{wilcoxon1992individual} to compare the pairwise transition weights in $\mathbold{W}_{f_{QL}}$ and $\mathbold{W}_{f_{LQ}}$ for each dataset.
Each cell in these matrices represents the knowledge transfer weight between a latent concept to another latent concept when the student transfers from one learning material type to another.
Since the weight values can have different ranges in the two weight matrices, we use a rank-based metric for their comparison.

Specifically, to compare $\mathbold{W}_{f_{QL}}$ and $\mathbold{W}_{f_{LQ}}$, we first flatten them and then run a Wilcoxon signed-rank test~\cite{wilcoxon1992individual} on them.
This test uses the median pairwise rank difference to identify if the rankings of between-concept transition weights are the same in the two matrices. %between  pairwise transition weights for each dataset.
With a small p-value of $0.02$ for the MORF dataset, we can conclude that the median pairwise difference between $\mathbold{W}_{f_{QL}}$ and $\mathbold{W}_{f_{LQ}}$ is non-zero, and these transition weights are significantly different in the MORF dataset.
This means that there are latent-concept pairs in MORF that can easily transfer to each other when the student transitions from assignments to video lectures (or video lectures to assignments), but they cannot transfer as easily when the students transition in a reverse order.
On the other hand, the p-values for EdNet and Junyi datasets are large ($>0.7$).
Therefore, we cannot reject the Null hypothesis and conclude that transition weights in $\mathbold{W}_{f_{QL}}$ and $\mathbold{W}_{f_{LQ}}$ are different in EdNet and Junyi.

% \begin{table}[!ht]
% \centering
% \caption{Spearman correlation coefficients with p-values}
% \resizebox{0.25\textwidth}{!}{
% \begin{tabular}{c|cc}
% \toprule
%       & \multicolumn{1}{l}{Correlation} & \multicolumn{1}{l}{p-value} \\
%                   \hline
% Junyi & 0.38443                    & 2.09e-37                    \\
% EdNet & 0.33680                   & 3.30e-08                    \\
% MORF  & -0.03686                  & 0.55714               \\
% \bottomrule
% \end{tabular}
% }
% \end{table}

As a second measure, we calculate the Spearman correlation coefficient~\cite{spearman1961proof} % and its associated p-values 
between the flattened $\mathbold{W}_{f_{QL}}$ and $\mathbold{W}_{f_{LQ}}$ (reported in Table~\ref{fig:spearman}).
As we can see, for EdNet and Junyi, $\mathbold{W}_{f_{QL}}$and $\mathbold{W}_{f_{LQ}}$ are positively correlated with a small p-value.
So, a higher (lower) transfer weight from questions to problem explanations in a specific latent concept pair in EdNet usually means a higher (lower) transfer weight from problem explanations to questions in the same concept pair.
But for MORF, the correlation coefficient is close to $0$. % and the p-value is large.
This is in accordance with our previous conclusion that the transition weight matrices are different in MORF. %, but more similar in EdNet and Junyi.

To further investigate, we visualize the weight matrices $\mathbold{W}_{f_{QL}}$ and $\mathbold{W}_{f_{LQ}}$.
For better visualization, we first perform a z-score normalization~\cite{patro2015normalization} on each weight matrix so that it has a mean of $0$ and a standard deviation of $1$. 
Then, we plot the heatmaps of these normalized transition matrices for all three datasets and show them in figure~\ref{fig:transition_analysis}.
As we can see, in the MORF dataset, the weight matrices are substantially different from each other, with a few similarities. 
For example, the values for latent dimensions row 8 to column 7 are close to each other ($1.4571$ in $\mathbold{W}_{f_{QL}}$ and $2.0010$ in $\mathbold{W}_{f_{LQ}}$). 
This indicates that the between-concept knowledge transfer from assignments to video lectures is different from the knowledge transfer from video lectures to assignments.

%This means that, although students forget similar knowledge from switching between assessed and non-assessed learning materials in general, there still are differences between knowledge forgetting from assessed to non-assessed materials and from non-assessed to assessed materials.
In contrast, the transition weights in $\mathbold{W}_{f_{QL}}$ and $\mathbold{W}_{f_{LQ}}$ are relatively similarly distributed in the Junyi and EdNet datasets.
However, there are still some small differences that can be observed. 
For example, the weights in Junyi are different from latent dimension row 7 to latent dimension row 14: the value is $2.0514$ in $\mathbold{W}_{f_{QL}}$, while it is $-2.5527$ in $\mathbold{W}_{f_{LQ}}$.
This shows that in Junyi and EdNet datasets, most concepts transfer similarly between different activity types. But for a few concept pairs, there are different transfer dynamics. 

These observations are consistent with the different dataset characteristics.
%as one possible reason for these different observations between MORF and EdNet or Junyi is the close-knit associations between different learning material types in Junyi and EdNet, unlike MORF.
Unlike MORF, in EdNet and Junyi there are close-knit associations between different learning material types.
For Junyi, the two types of materials are problems and hints, and for EdNet they are questions and problem explanations.
Each hint in Junyi (and each problem explanation in EdNet) is designed to help a single problem (or question) associated with it.
As a result, the transitions usually happen between similar learning materials with related concepts.
But, in MORF each assignment includes multiple problems and covers many concepts. 
Similarly, the video lectures are more general than Junyi's hints and EdNet's explanations and introduce a wider range of topics.
As a result, the students can transition between diverse and unrelated concepts. 
This leads to a more complicated association between MORF's assignments and video lectures, which in turn leads to more complex and dissimilar weight matrices $\mathbold{W}_{f_{QL}}$ and $\mathbold{W}_{f_{LQ}}$.

% concepts are atomic and simple. so they learn it by memorization and can transfer in any direction,

%This analysis demonstrates the need for knowledge transfer modeling in different learning activity types.
This analysis shows that knowledge transfer weights could depend on the transition order (permutation) between material types, especially for the datasets in which assessed and non-assessed material types are more complex and are not closely associated with each other.
Also, this analysis shows how to interpret knowledge transfer between different learning materials. 
This could help instructors in arranging course learning materials for the maximum possible knowledge transfer.

\subsection{Student Knowledge State Visualization}
\label{sec:experiments:knowledge}
\begin{figure*}[!htbp]
    \centering
    \includegraphics[width=0.85\textwidth]{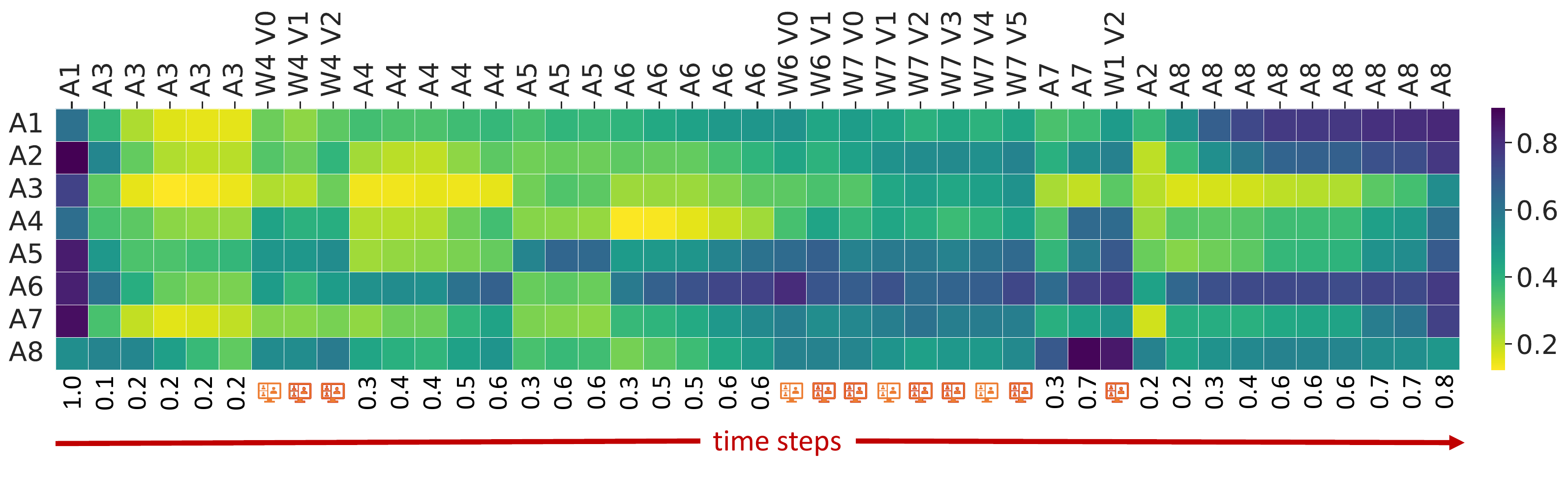}
    % \vspace{-10pt}
    % \caption{Visualization of student's knowledge (estimated performance) at each time step from the MORF dataset. 
    % Top x-ticks labels are learning source titles that student interacted with at each time step, bottom x-ticks labels are student performance or the 'book' icon for nun learning material activities, and y-ticks labels are assessed learning materials titles.}
    % \caption{Student's knowledge visualization of a sample student from MORF dataset. 
    % Top x-axis ticks are learning material titles that the student has tried at each time step, bottom x-axis ticks are real student performance or the 'screen' icon for videos. y-axis ticks are assignment titles. Each cell denotes the student's predicted performance in an assignment (shown by y-axis ticks) at each time step.
    % }
    \caption{Visualization of predicted student performance (as a knowledge indicator) for a sample student in the MORF dataset. The top x-axis ticks are learning material titles the student has tried at each time step. The bottom x-axis ticks are real student performance (in assessed activities) or the ‘screen’ icon (in non-assessed ones). The y-axis ticks are the assessed material titles. Each cell shows the student’s predicted performance in an assessed material at each step.}
    \label{fig:knowledge_state}
    \vspace{-10pt}
\end{figure*}

% 2) In Fig 3, the author said “Student’s knowledge visualization” but then it is also said that “calculate the student’s predicted performance p t+1 in each assessed learning material, as an indicator of student knowledge”. It makes me confuse whether this visualization is from the weights of a hidden layer or just predicted outcomes. 
% update the caption to include performance as an indicator ...

%Another important experiment is to evaluate whether our TAMKOT can track interpretable student knowledge states. 
To study student knowledge state interpretation, we visualize the student mastery level of each assessed learning material at each time step.
%We treat the predicted student performance of each assessed learning material as the student knowledge state after each interaction.
In other words, after every activity, we use equation\ref{eq: prediction} to calculate the student's predicted performance $p_{t+1}$ in each assessed learning material, as an indicator of student knowledge.
% For this case study, we sample one student's trajectory from the MORF dataset to visualize (Figure~\ref{fig:knowledge_state}).
For this case study, we base our analysis on the MORF dataset. 
As Junyi and EdNet have similar patterns as MORF, we don't show their visualization due to the page limitation. 
We sample one student's trajectory from the MORF dataset to visualize (Figure~\ref{fig:knowledge_state}).
Each row shows the student's predicted performance in one of the eight course assignments during the learning trajectory.
Each column shows the student's predicted performance in all assignments after attempting a particular learning activity.
The title of the attempted learning activity is shown at the top of each column.
If the student has attempted a problem, their observed performance is shown at the bottom of each column.
If the student has watched a video lecture, a 'screen' icon is shown at the bottom of each column.
%The student's performance for the corresponding interaction or the 'video' icon for the video lecture activities.
%Top x-ticks represent the title of the learning material that the student interacted with at each time step.
% For better visualization, we use ``W* V*'' as an abbreviation of video * of week *, and ``A*'' to summarize Assignment *. 
We abbreviate video lecture * of week ** as ``W** V*'', and Assignment * as and ``A*''.
So, `` W4 V1'' represent the first video of week 4. % and ``A1'' represent Assignment 1.
%The ticks on the horizontal axis are either 
% y-ticks are labeled as the assessed learning material titles that the student's knowledge (performance) is estimated for them.
%The ticks on the vertical axis are labeled as the assignments titles. 
%Each cell denotes the predicted performance of an assignment after the student's activity at each time point.

We see the student's initial estimated knowledge is high, after the first interaction. 
This happens since the student received the full score of $1.0$ in the first assignment. %, our TAMKOT learns this student may have high initial knowledge. 
But, as we observe 5 low-grade attempts of A3, the student's mastery level drops. % then dropped when we observing 5 low-grade attempts of A3. 
This could be because the student skipped A2 and video lectures for week 2.
We then see the student's knowledge grows by watching video lectures of week 4.
% However, the knowledge improvement value for each assignment from watching different video lectures is different.
However, \textit{watching different videos produce different knowledge improvement values for each assignment.}
For example, after watching week 4's lectures, although the student's mastery level of each assignment increased, A4 is the one that has the largest improvement.
%Then, we notice that the student's score of A4 increases, and the student's knowledge for A4 increases as well. 
% We observe similar increases of student knowledge for the student's activities of A5, A6, and A8: the knowledge for corresponding assignment increases when the score increases.
We also observe an \textit{increase in knowledge as the score of the corresponding assignment increases.}
For example, as student scores increases, their knowledge of A5, A6, and A8 also increases.
Moreover, when the student watches the multiple lectures between two assignments, the first attempt usually has the largest improvement.
It means that \textit{student knowledge does not keep growing while continuously watching multiple lectures one after another.}
For example, the student's knowledge grows significantly after watching W4 V0, but the next two attempts of watching W4 V1 and W4 V2 only improve the student's knowledge slightly.
This conclusion is in line with previous research that shows assessed activities could be more helpful than repeating non-assessed ones~\cite{koedinger2015learning,mirzaei2020structure}.
% This conclusion is in line with previous research that shows assessed activities could be more helpful than repeating non-assessed ones, especially for students with a relatively high initial knowledge~\cite{koedinger2015learning,mirzaei2020structure}.

% \subsection{Learning Material Concept Clustering}
% \label{sec:experiments:material}
% \input{5.7-resource.tex}

\section{Conclusions and Future Work}
\label{sec:conclusions}
% In this paper, we propose a multi-type deep transitional knowledge tracing model (TAMKOT), which can explicitly model student knowledge transfer between multiple types of learning materials. 
% The proposed method learns multiple knowledge transfer matrices, and models different learning material types in different embedding space.
% The extensive experiments on three real-world datasets show the TAMKOT can get similar, or better results than various state-of-art baselines. The student knowledge state visualization and learning material concept clustering indicate that TAMKOT learns meaningful knowledge states and provides the interpretable latent learning material concepts.
% Our analysis on knowledge transfer between different learning material types demonstrates that the knowledge transfer for different permutations of learning material types is different.

In this paper, we proposed Transition-Aware Multi-Activity Knowledge Tracing (TAMKOT), to model student learning from both assessed and non-assessed learning activities and explicitly learn the knowledge transfer between different learning activity types.
TAMKOT learns multiple knowledge transfer matrices, one for each transition type between student activities, and allows for unlimited transitions between learning activity types in any order.
%different latent spaces for representing each material type.
We performed extensive experiments on three real-world datasets and compared TAMKOT with state-of-the-art baselines in predicting student performance.
We also analyzed and interpreted the learned knowledge transfer matrices and student knowledge states.
Our experiment results showed that explicitly modeling both assessed and non-assessed activities in TAMKOT, in addition to the transition-aware knowledge transfers between them, is necessary to accurately represent student knowledge and predict their performance.
We also concluded that the amount of knowledge transfer between concepts could depend on the transition order (permutation) between activity types, especially for the datasets in which assessed and non-assessed material types are more complex.
Finally, we showcased a sample student's knowledge states and their interpretation that for that particular student, the assessed activities were more helpful than the non-assessed ones.
In the future, we would like to explore TAMKOT's performance in supporting more than two learning activity types and investigate the knowledge transfer among them.

% \section{Appendix}
% \vspace{-3pt}
% \label{sec:conclusions}
% \input{7-appendix.tex}

\subsection*{Acknowledgments}
This paper is based upon work supported by the National Science Foundation under Grant No. 2047500.

%\subsection{References}
% \clearpage

\bibliographystyle{IEEEtran}
\bibliography{refs}

\end{document}